\documentclass[apjl]{emulateapj}

\newcommand{\OIII}{[\mbox{O\,\textsc{iii}}]}

\newcommand{\NII}{\mbox{N\,\textsc{ii}}}

\newcommand{\kms}{km s$^{-1}$}
\newcommand{\Ha}{H$\alpha$}   
  
\newcommand{\msigma}{M$_{\rm BH}$-$\sigma_*$}
\newcommand{\msun}{M$_{\odot}$}
\newcommand{\VOIII}{V$_{\rm OIII}$}
\newcommand{\SOIII}{$\sigma_{\rm OIII}$  }
\newcommand{\SVD}{$\sigma_{*}$}

\shorttitle{The prevalence of gas outflows in Type 2 AGNs. III}
\shortauthors{Woo et al.}

\begin{document}

\title{Delayed or no feedback? - Gas outflows in Type 2 AGNs. III.}
\author{Jong-Hak Woo$^{1}$}
\author{Donghoon Son$^{1,}$}
\author{Hyun-Jin Bae$^{1,2}$}

\affil{$^{1}$Astronomy Program, Department of Physics and Astronomy, Seoul National University, Seoul 151-742, Republic of Korea; woo@astro.snu.ac.kr}
\affil{$^{2}$Department of Astronomy and Center for Galaxy Evolution Research, Yonsei University, Seoul 120-749, Republic of Korea; hjbae@galaxy.yonsei.ac.kr}

\begin{abstract}
We present gas kinematics based on the \OIII $\lambda$5007 line and their connection to galaxy gravitational potential,
active galactic nucleus (AGN) energetics, and star formation, using a large sample of $\sim$110,000 AGNs and star-forming (SF) galaxies at z$<$0.3. 
Gas and stellar velocity dispersions are comparable to each other in SF galaxies, indicating that the ionized gas kinematics can be accounted by the 
gravitational potential of host galaxies. In contrast, AGNs clearly show non-gravitational kinematics, which is comparable to or stronger than the virial motion caused by the 
gravitational potential. The \OIII\ velocity-velocity dispersion (VVD) diagram dramatically expands toward high values as a function of AGN luminosity,
implying that the outflows are AGN-driven, while SF galaxies do not show such a trend.  
We find that the fraction of AGNs with a signature of outflow kinematics, steeply increases with AGN luminosity and Eddington ratio. In particular, the majority of luminous AGNs presents strong non-gravitational kinematics in the \OIII\ profile.
AGNs with strong outflow signatures show on average similar specific star formation rate (SSFR) 
to that of starforming galaxies. In contrast, AGNs with weak or no outflows have an order of magnitude lower SSFR,
suggesting that AGNs with current strong outflows do now show any negative AGN feedback and that it may take the order of a dynamical time 
to impact on star formation over galactic scales.
\end{abstract}


\section{Introduction}

Active galactic nuclei (AGNs) are often believed to play a significant role
in galaxy evolution. As the observed scaling relations between black hole mass and global properties of their host galaxies seem to require self-regulation
between black hole growth and galaxy evolution \citep{Kormendy&Ho2013}, theoretical models often include AGN feedback 
as a core component of galaxy evolution \citep[e.g.,][]{Dimatteo+05,Croton06,Dubois+13,deGraf+15, Dubois+16,Hopkins+16}. Nonetheless, understanding how black holes and galaxies coevolve over Hubble time is observationally limited due to the scarcity of direct measurements of scaling relations at high redshift \citep{Woo+06,Treu+07,Woo+08,Bennert+11,Park+15}, and 
the nature of AGN feedback and the coupling of the AGN energy output with the interstellar matter (ISM) is a subject of active research in this field.

AGN gas outflows are often observed over galaxy scales \citep{Nesvadba+06, Liu+13, Harrison+14, Husemann+14, Husemann+16, Karouzos+16a}, hence, outflows may function as an effective channel of  AGN feedback. 
By connecting nuclear activity with the larger scale ISM and star formation (SF), AGN outflows may push gas out of host galaxies and/or heat up the ISM, leading to
suppression of SF \citep{Silk+98, Fabian12}. On the other hand, outflows may trigger SF by compressing the ISM \citep{Zubovas+13, Ishibashi+14}.
Currently, the evidence of negative AGN feedback on SF is yet to be conclusive \citep[e.g.,][]{Vilar-Martin+16}, 
while detections of positive feedback on SF have been reported by several observational works based on individual objects \citep[e.g.,][]{Cresci+15, Carniani+16}.  Thus, it is of importance to
investigate the true nature of AGN feedback and understand the overall effect of AGNs through gas outflows. 

The direct connection between AGN activity and SF seems complex as various observations have been presented in the literature. 
For typical AGNs, a positive correlation has been reported between AGN luminosity, probed by either optical or X-ray, and SF luminosity \citep[e.g.,][]{Netzer09,Diamond-Stanic+12,Woo+12,Matsuoka15}, indicating an average scaling between AGN activity and SFR. 
In contrast, at high redshift in particular, X-ray AGNs seem to show enhanced SFR \citep{Rosario+12,Santini+12}. 
Since the SFR of AGN host galaxies is difficult to measure and there may be systematic differecne among various SFR indicators
based on UV, optical, IR signatures \citep[e.g.][]{Matsuoka15, Rosario+16}, more detailed investigation is required to reveal the nature of AGN and SF connection. 
In comparing with non-AGN galaxies, a recent study by \citet{Shimizu+15} reported that X-ray and optical AGNs
tend to have lower specific star formation rate (SSFR) compared to star-forming galaxies. A similar work by \citet{Ellison+16b} showed 
that radio and optical AGNs have on average lower SSFR than star-forming galaxies, while mid-infrared selected AGNs show comparable or slightly higher SSFR than star-forming galaxies.  These results based on the present-day galaxies may suggest that AGN feedback suppresses SF. However, the physical mechanism 
for connecting AGN and SF is yet to be clear, requiring further detailed studies of the link between outflows and SF. 

To understand the role of AGN gas outflows in galaxy evolution, it is important to investigate how common and strong these outflows are so that AGNs as a population, instead of individual objects, can be used for connecting outflows with AGN energetics and galaxy evolution. 
While the demography of ionized gas outflows in AGNs was limited to relatively small samples in the past \citep{Nelson&Whittle96, bo05, greene&ho05}, more robust results became available recently based on statistical samples of AGNs from large surveys \citep[e.g.,][]{mullaney+13, Bae&Woo14, Woo+16}. 

To perform a statistical investigation of gas outflows and their connection to AGN activity, we are performing a series of studies using a large sample of type 2 AGNs from Sloan Digital Sky Survey (SDSS). In the first of this series, \citet[][here after, Paper 1]{Woo+16} presented a census of ionized gas outflows based on the \OIII$\lambda$5007 kinematics,
by analyzing a sample of 39,000 present-day type 2 AGNs. At least 50\% of AGNs over the full luminosity range explored in that study, shows kinematic signatures of gas outflows.
This outflow fraction is considered as a lower limit for low-luminosity AGNs since their outflow indicator (i.e., a wing component of the \OIII\ line) is relatively weak and the gravitational potential of their host galaxy dominates in forming the emission line profile. In fact, for high-luminosity AGNs, the outflow fraction is over 90\%, indicating that 
the majority of energetic AGNs show outflows. The steep increase of the outflow fraction with Eddington ratio indicates a close link between outflow kinematics and AGN accretion. In contrast, we find no strong correlation between radio luminosity and ionized gas kinematics, indicating that the outflows are not directly 
connected to radio activity for most AGNs. In the second of this series, \citet{Bae&Woo16} presented 3-D biconical outflow models
combined with a thin dust plane, in order to constrain the intrinsic physical nature of gas outflows. Our models successfully reproduced the observed emission line profiles
and the distribution of the velocity-velocity dispersion (VVD) diagram measured from the \OIII\ line. Based on the Monte Carlo simulations of the VVD distribution, we constrained the physical parameters
of the outflows, including launching velocity, opening angle, and dust extinction \citep[][hereafter Paper II]{Bae&Woo16}.

In this paper, we mainly compare AGNs and SF galaxies in terms of the \OIII\ kinematics and their link to SF, in order to investigate the role of gas outflows as AGN feedback.
Throughout the paper, we use the cosmological parameters as
$H_0 = 70$~km~s$^{-1}$~Mpc$^{-1}$, $\Omega_{\rm m} = 0.30$, and $\Omega_{\Lambda} = 0.70$.

\section{Sample \& Analysis}

\subsection{Sample selection}

\begin{figure} 
\centering
\includegraphics[width=0.38\textwidth]{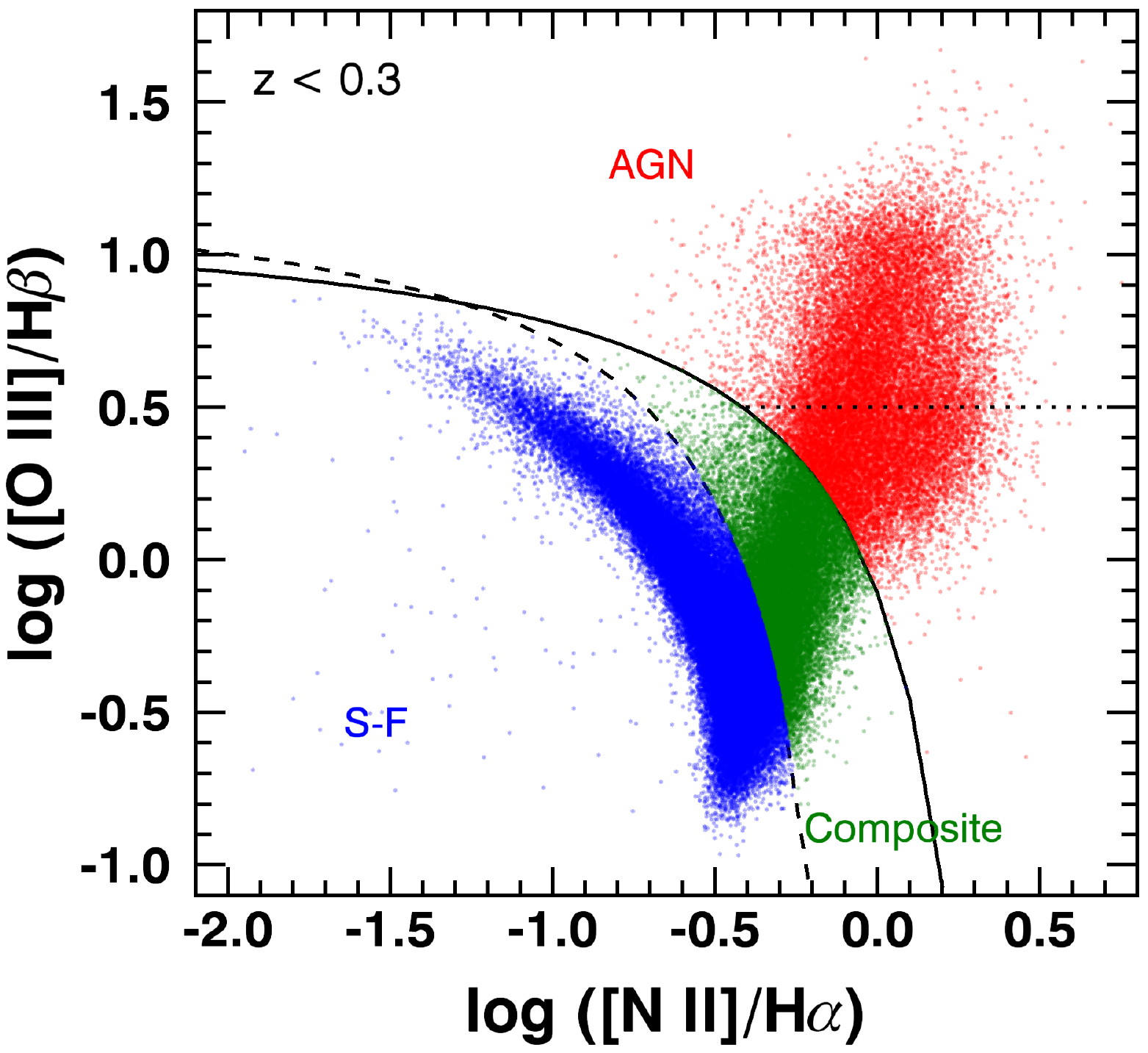}
\caption{Classification of $\sim$110,000 galaxies at z$<$0.3, with well-defined emission-lines.
}
\end{figure}

We extend our previous sample of AGNs in Paper I to all emission line galaxies at z$<$$\sim$0.3, by including star-forming galaxies.
The initial sample of 235,922 emission line galaxies is selected from the SDSS Data Release 7 \citep{abazajian+09}, using the MPA-JHU Catalogue\footnote{\url{http://www.mpa-garching.mpg.de/SDSS/}}, with signal-to-noise ratio (S/N) $\ge$ 10 in the continuum, and S/N $\ge$ 3 
in the four emission lines, H$\beta$, \OIII $\lambda$5007, \Ha, and [\NII] $\lambda$6584, which are used to classify each object 
in the diagnostic diagram (see Fig. 1). Using this initial sample, we applied a further constraint to select
objects with well-defined emission line profiles with the amplitude-to-noise (A/N) ratio larger than 5
for the \OIII\ and \Ha\ lines. As a result, we obtained a total of 112,726 emission line galaxies, which are classified into three
groups: AGNs, composite objects, and star-forming galaxies as listed in Table 1, based on the classification scheme by \citet{ka03}. 
In the first paper of this series, we presented the detailed analysis on the pure AGN and composite samples (Paper I). In this work,
we mainly compare AGNs with SF galaxies in terms of \OIII\ kinematics, and investigate how gas outflows are related to star-formation. 

\begin{table}
\begin{center}
\caption{Number of objects in each group}
\begin{tabular}{cccc}
\tableline\tableline
redshift & AGN & Composite  & S-F \\
(1)&(2)&(3)&(4)\\
\tableline
0.003 $<$z$<$ 0.1  & 14923 (6733) & 14714 (4700) & 58928 (23999)\\ 
0.1 $<$z$<$ 0.2   & 7382 (4495) & 5614 (1857) & 10019 (2967) \\
0.2 $<$z$<$ 0.3   & 447 (289) & 384 (147) & 315 (86)  \\
0.003 $<$z$<$ 0.3  & 22752 (11517) & 20709 (6704) & 69265 (27052)\\
\tableline
\end{tabular}
\tablecomments{(1) Redshift range; (2) Number of objects in AGN group; (3) Number of objects in Composite group;
(4) Number of objects in SF group. The numbers in parenthesis indicate the number of objects with
a double Gaussian \OIII\ profile.}
\end{center}
\end{table}

\subsection{Sample properties}

We derive the properties of each galaxy in the sample in order to compare them with ionized gas kinematics, as we presented in detail for AGNs in Paper 1.
Here, we briefly summarize sample properties. 
First, we adopt the stellar velocity dispersion and stellar mass from the MPA-JHU catalogue to investigate the
host galaxy mass scale and gravitational potential.
In the following analysis, we include stellar velocity dispersion ($\sigma_{*}$) value down to 30 \kms, which is slightly lower than the instrumental resolution of the SDSS spectra while we exclude unreliably values ($\sigma_{*}$ $>$ 420 \kms) as recommended by the SDSS catalogue\footnote {http://classic.sdss.org/dr7/algorithms/veldisp.html}. 
Since stellar velocity dispersions are more uncertain for lower mass galaxies due to the spectral resolution limit as well as the lower S/N,
we will also use the stellar mass (M$_{*}$) estimates as a tracer of the host galaxy gravitational potential. 
As shown in Fig 2, the sample galaxies follow the correlation between stellar mass and stellar velocity dispersion as expected from
Fiber-Jackson relation \citep{Faber&Jackson76} although the scatter becomes larger for lower mass galaxies due to measurement uncertainties.
Note that three groups have different mass ranges. While the stellar mass of the most AGN host galaxies ranges from 10$^{9.5}$ to 10$^{11.5}$ \msun,
that of SF galaxies lies between $\sim$10$^{8}$ to 10$^{11}$ \msun. Thus, when we compare AGNs with SF galaxies, we will use stellar mass fixed subsamples,
to account for the difference of their gravitational potential. 

In addition to the host galaxy properties, we also determine the properties of black holes. 
For black hole mass estimates, we use the updated scaling relations, e.g., black hole mass-stellar velocity dispersion (\msigma) relation based on the reverberation-mapped AGNs from \citet{woo+15} \citep[see also][]{Woo+10,park+12,Grier+13}
and black hole mass - stellar mass relation \citep{marconi&hunt03}. As we pointed out in Paper I, the range of black hole mass significantly changes
depending on the adopted scaling relation. We adopted the black hole mass estimates based on the black hole mass - stellar mass relation to be consistent with Paper 1, while the choice of the black hole mass estimates do not change the main results presented in this paper. 

\begin{figure} 
\centering
\includegraphics[width=0.48\textwidth]{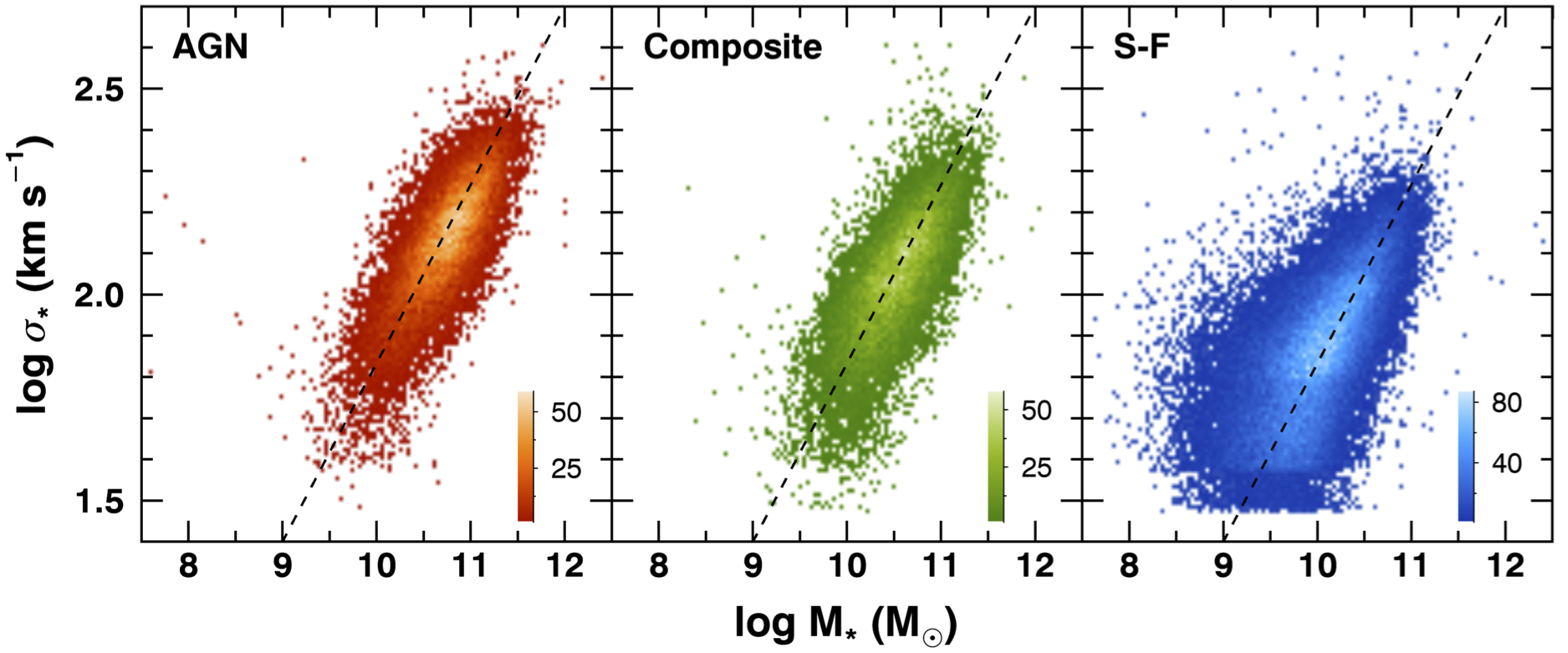}
\caption{Stellar velocity dispersion vs. stellar mass, respectively for AGNs (left), composite objects (middle), and SF galaxies (right).
The color bar represents the number density calculated within a small bin with $\Delta$x= 0.04 dex and $\Delta$y=0.01 dex.
}
\end{figure}

\begin{figure*}
\centering
\includegraphics[width=.86\textwidth]{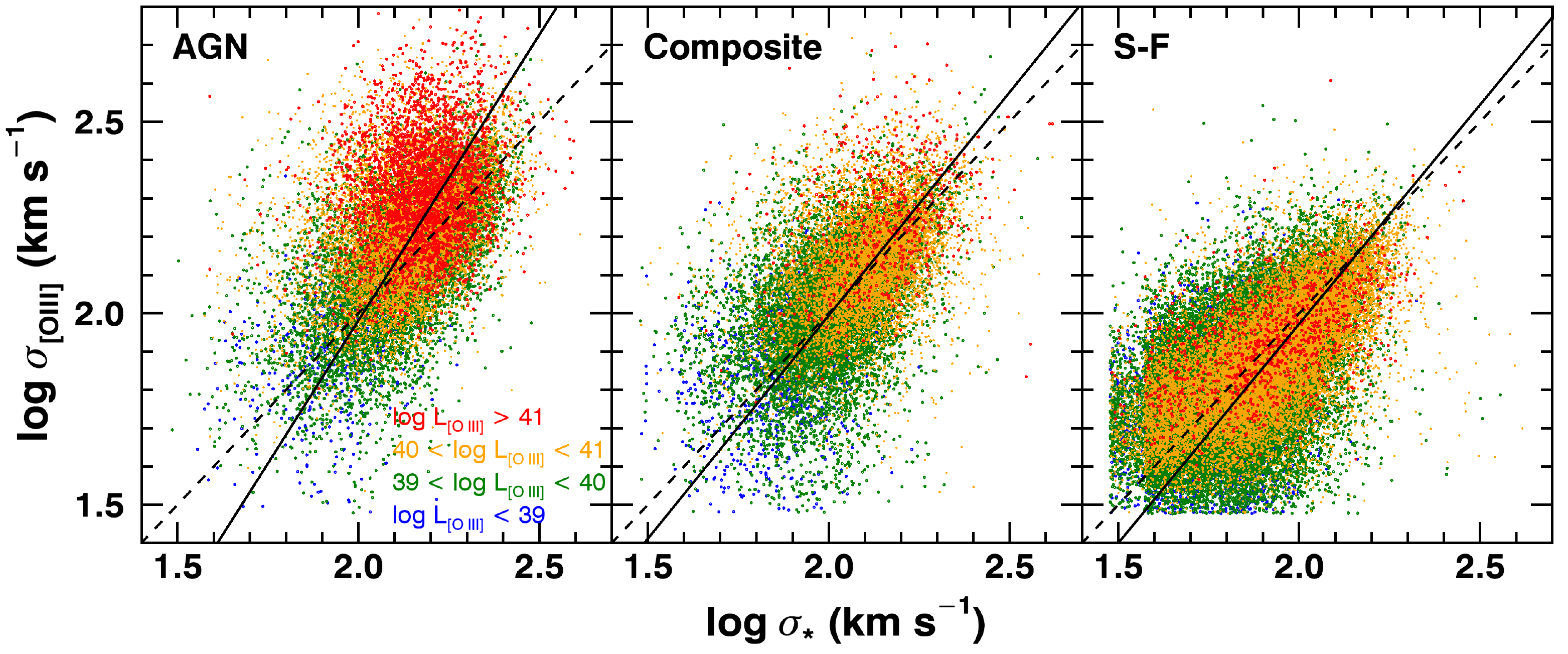} 
\caption{Comparison of the \OIII\ gas and stellar velocity dispersion for AGNs (left),
composite objects (middle), and SF galaxies (right). Colors represent the \OIII\ luminosity range of each object. 
The best-fit slop (black solid line; 1.49 for AGNs, 1.17 for composite objects, 1.14 for SF galaxies) indicates 
that the relation is not linear for AGNs while composite objects and SF galaxies show consistency between gas and stellar kinematics.}
\end{figure*}

\subsection{\OIII\ kinematics}

We used the strong \OIII\ line at 5007\AA\ line as a tracer of the ionized gas outflows, by measuring the velocity shift from the systemic velocity
and the velocity dispersion compared to the stellar velocity dispersion, as we discussed in Paper I.
Here we briefly summarize the procedure for completeness.
Using the SDSS spectra, we first fitted stellar absorption lines with a series of stellar population models, which were constructed with the penalized pixel-fitting code \citep{ce04} based on simple stellar population models with solar metallicity and various ages (60 Myrs to 12.6 Gyrs) \citep{sb06}. 
Based on the stellar absorption fit, we determined the systemic velocity, which will be used for calculating the velocity shift of
\OIII. 

By subtracting the best-fit stellar population model, we generated pure emission line spectra, on which 
we fitted the \OIII\ line with one or two Gaussian components using the MPFIT code (Markwardt 2009). 
Only if  the \OIII\ line profile shows a prominent wing component, i.e., the peak of the second component is a factor of three larger than the noise level (i.e., A/N ratio $>$ 3), we adopted the 2nd Gaussian component, in order to avoid unreliable detection of wing features. 

Once we obtained the best fit model for \OIII, we calculated the first and second moments of
the line profile using the best-fit model as
\begin{eqnarray}
\lambda_{0} = {\int \lambda f_\lambda d\lambda \over \int f_\lambda d\lambda}.
\end{eqnarray}
\begin{eqnarray}
[\Delta\lambda_{\OIII}]^2  = {\int \lambda^2 f_\lambda d\lambda \over \int f_\lambda d\lambda} - \lambda_0^2, 
\end{eqnarray}
where $f_\lambda$ is the flux at each wavelength.
Then, we calculated the velocity shift (\VOIII) of the first moment of \OIII\ with respect to the systemic velocity, which is determined from stellar absorption lines, and the velocity dispersion (\SOIII) of \OIII\ from the second moment. 
We determined the uncertainty of \VOIII\ and \SOIII\ by performing Monte Carlo simulations, which generate the distribution of measurements based on the 100 mock spectra generated by randomizing flux with flux error at each wavelength. 
We adopted the 1 $\sigma$ dispersion of the distribution as the measurement uncertainty, respectively for \VOIII\ and \SOIII.

When we examined the fractional error of $\sigma_{\OIII}$ of the sample, the distribution shows
a Gaussian distribution with a mean value of  $-0.80\pm0.50$ in log scale (-0.89$\pm$0.44 for AGNs, -0.74$\pm$0.39 for composite objects, and -0.79$\pm$0.54 for SF galaxies), which corresponds to $\sim$16\% uncertainty. 
To avoid uncertain measurements, we excluded 7902 objects, that have a large uncertainty (i.e. fractional error $>$ 1), from the sample (7.0\%) when we used the measurements of the \OIII\ kinematics in the analysis. 
In the case of \VOIII, the mean uncertainty is $29.1\pm19.2$ \kms\ for AGNs, $30.8\pm54.8$ \kms\ for composite objects, and $23.9\pm16.3$ \kms\ for SF galaxies, indicating that very weak velocity shifts (i.e., $<$ 30 \kms) are difficult to detect. Note that the quoted errors are the r.m.s from the distribution of errors of individual objects.
The measured \OIII\ velocity dispersions were corrected for the wavelength-dependent instrumental resolution of the SDSS spectroscopy, which is close to 55-60 \kms\ at the observed wavelength of \OIII. A number of objects has relatively narrow \OIII, for which the measured $\sigma_{\OIII}$ is smaller than the instrumental resolution. We excluded 1872 objects with very low  velocity dispersion (i.e., $\sigma_{\OIII}$ $<$ 30 \kms) or unresolved \OIII.
In total we excluded 8.7\% of the sample, which have either very small velocity dispersion (i.e., $\sigma_{\OIII}$ $<$ 30 \kms) or large fractional errors ($>$1). Note that including these objects does not change our conclusions in the following analysis.

\section{Results}

\begin{figure}
\centering
\includegraphics[width=.45\textwidth]{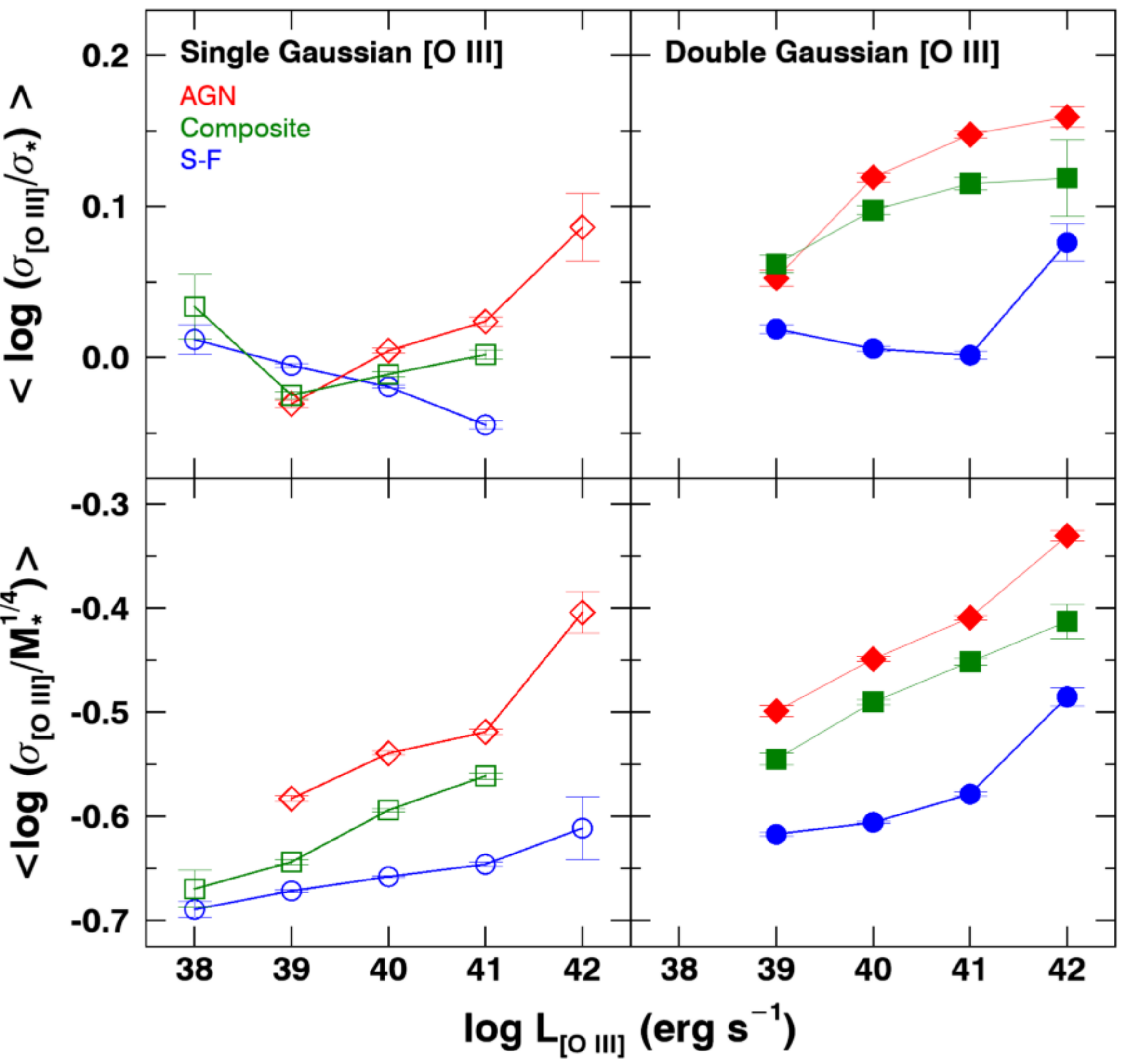} \\
\caption{Mean ratio of \OIII-to-stellar velocity dispersions as a function of \OIII\ luminosity (top),
respectively for AGNs (red), composite objects (green), and SF galaxies (blue).
We divide the sample into two groups: \OIII\ fitted with single Gaussian (left) and double Gaussian (right).
The ratio of the AGN sample is systematically higher than that of SF galaxies, 
indicating that the non-gravitational component, i.e., outflows is dominant in AGNs, particularly at high luminosity regime. 
The bottom panels show the mean ratio of \OIII\ divided by stellar mass (in units of \kms\ M$_{\odot}^{-1}$) as a function of \OIII\ luminosity. 
}
\end{figure}

\begin{figure}
\centering
\includegraphics[width=0.45\textwidth]{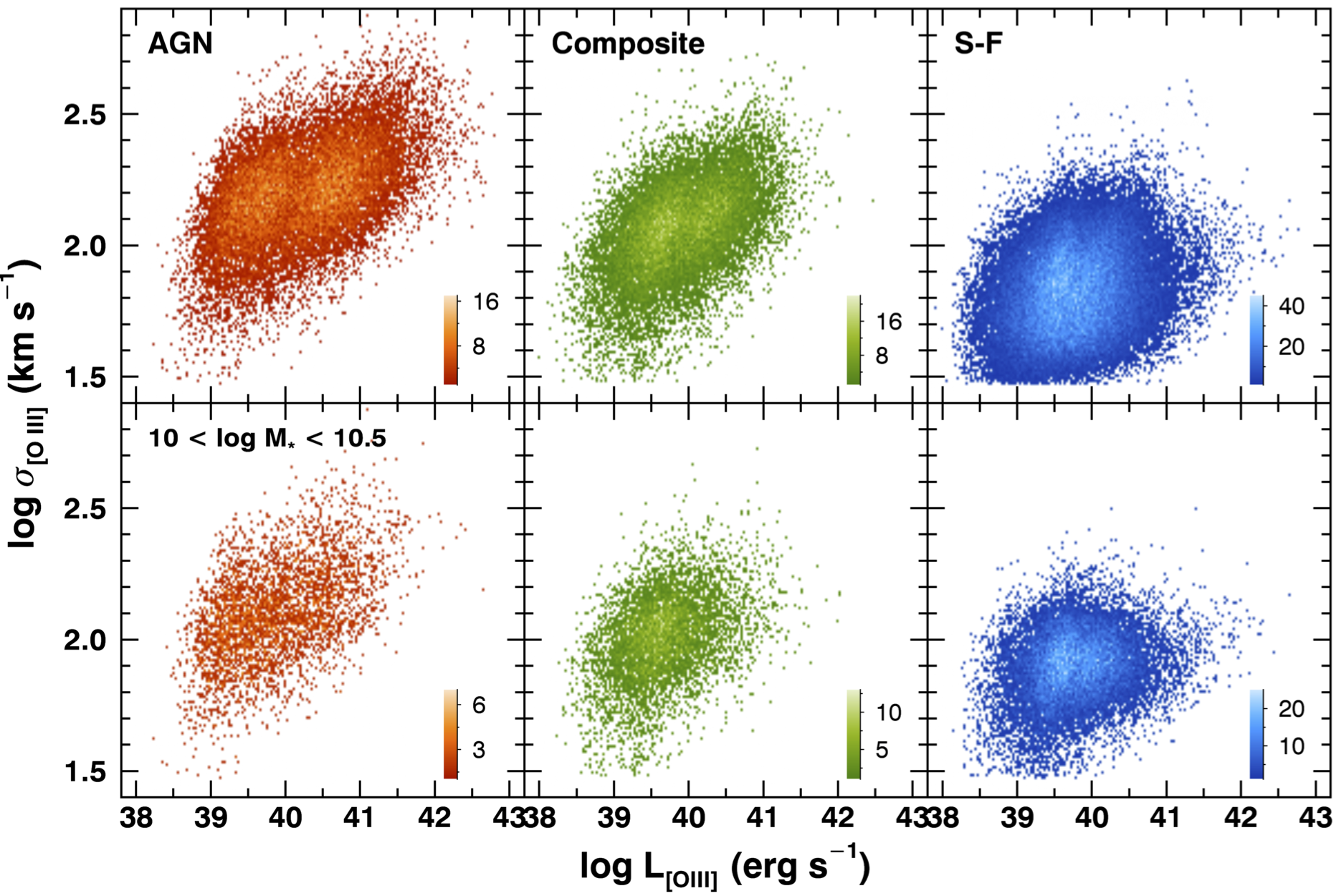}
\caption{Comparison of \OIII\ velocity dispersion with \OIII\ luminosity, respectively for AGNs (left), composite objects (middle), and SF galaxies (right),
using the entire sample (top) and for the stellar mass-limited sample (bottom).
The color represent the number density in each bin with $\Delta$x= 0.03 dex and $\Delta$y=0.01 dex.}
\end{figure}

\subsection{gravitational vs. non-gravitational kinematics}

To investigate the signature of gas outflows, we compare gas and stellar kinematics using \OIII\ velocity dispersion (\SOIII) and stellar velocity dispersion (\SVD) in Figure 3. 
We find a clear difference between AGN sample and SF galaxies. AGNs with larger \SOIII\ have larger \SVD,
indicating that the host galaxy gravitational potential plays a role in broadening the gas emission lines.
However, the relation is not linear as \SOIII\ is larger than \SVD\ particularly for high luminosity AGNs.
As we discussed in Paper I, we interpret the ratio of \SOIII\ to \SVD\ larger than unity as an indication of non-gravitational kinematic components, i.e. outflows. 

In contrast, SF galaxies show almost one-to-one relationship between \SOIII\ and \SVD, suggesting that
the broadening of gas emission line can be entirely accounted by the virial motion due to the galaxy gravitational potential. 
Note that the non-linear relation between \SOIII\ and \SVD\ for the AGN sample is mainly due to the wing component in the \OIII\ line profile,
which is not well constrained when FWHM is used to represent the velocity of \OIII\ line (see Paper I for detailed discussion). 
In the case of composite objects, \SOIII\ and \SVD\ are linearly correlated each other. 
Although AGN outflows may be present in composite objects, the outflow velocity is relatively low due to the average low luminosity (see the lack of high \OIII\ velocity AGNs in Figure 3), hence the non-gravitational kinematic component can be easily diluted by the dominant virial component. 

In Figure 4, we investigate the ratio between \SOIII\ and \SVD\  as a function of \OIII\ luminosity. Here we separate the emission line galaxies, depending on the presence of a wing component in \OIII\ (i.e., single Gaussian vs. double Gaussian). 
Regardless of AGN activity, most objects without a wing component show the \SOIII/\SVD\ ratio close to unity except for the AGNs in the highest luminosity bin,
indicating that \OIII\ gas kinematics of these objects are mostly due to the host galaxy gravitational potential. 

In contrast, among the objects with a wing component in \OIII\ (right panels in Figure 4), there is a systemic difference of the \SOIII/\SVD\ ratio between AGNs and SF galaxies. 
For AGNs, \SOIII\ is much larger than \SVD\ (by 0.13 dex on average), indicating the presence of non-virial kinematics, while for SF galaxies, \SOIII\ is comparable to \SVD\ except for the highest luminosity bin (the average ratio log (\SOIII/\SVD) = 0.01). Note that we detect a wing component in \OIII, which is fitted by the second Gaussian component, for $\sim$39\% of SF galaxies. However, the \SOIII\ values are similar to stellar velocity dispersions. 

Instead of \SVD, we can use stellar mass to normalize gas outflow kinematics by the host galaxy gravitational potential (bottom panels of Figure 4).
We find qualitatively same trends that the broadening of \OIII\ in AGNs cannot be entirely accounted by the gravitational potential
while the broadening of \OIII\ in SF galaxies are mainly due to the host galaxy potential. The average difference between AGNs and SF galaxies is 0.16 dex. Note that even among the objects with single Gaussian \OIII, we clearly detect the systematic 
difference of the ratio between AGNs and SF galaxies (the average difference is 0.11 dex).

\subsection{\OIII\ velocity dispersion vs. luminosity}

We investigate how the \OIII\ kinematics are related to \OIII\ luminosity 
for AGNs and SF galaxies in Figure 5. 
There is an increasing trend of \SOIII\ with increasing luminosity in AGNs although the trend is very broad. 
The increasing trend is also present among composite objects while the the luminosity range of \OIII\ is an order of magnitude narrower.
These trends show that gas outflows are stronger in AGNs with higher luminosity, indicating AGN energetics are responsible for the outflow kinematics. 
In contrast, we find no clear increasing trend among SF galaxies. On average the \SOIII\ is lower in SF galaxies while the range of \SOIII\ is as broad as that of AGNs.
Since the average stellar mass of AGNs and SF galaxies is very different as we pointed out in S.2.2,  we plot the same comparison after limiting stellar mass
between 10$^{10}$ and 10$^{10.5}$ (bottom panels in Fig. 5). Again, we find that AGNs and SF galaxies show different trends, suggesting that  
these different trends are not due to the difference in mass scales between AGN and SF galaxy samples.

\subsection{\OIII\ velocity-velocity dispersion diagram}

In Figure 6, we combine the velocity and velocity dispersion of \OIII\ to investigate the difference of gas kinematics between AGNs and SF galaxies. We find a dramatic contrast between AGNs and SF galaxies in terms of the range of gas velocities. 
While SF galaxies are concentrated at the lower center with a limited range of velocity shift ($|$\VOIII$|$ $<$ 100 \kms) and velocity dispersion (\SOIII\ $<$ $\sim$150 \kms), AGNs show a much larger distribution, including AGNs with extreme kinematics (i.e., $|$\VOIII$|$ $>$ 200 \kms\ and \SOIII$> $400 \kms).

The average \SOIII\ is $175.1\pm76$, $125.3\pm49.9$,
and $73.2\pm28.6$ \kms, respectively for AGNs, composite objects, and SF galaxies, indicating that AGNs have more than a factor
of 2 broader \OIII\ than SF galaxies. Also, the dispersion of the \SOIII\ distribution is significantly different with a factor of $\sim$3 broader distribution in AGNs. Note that the \SOIII\ distribution is skewed toward high values, indicating the presence of extreme kinematics in the AGN sample. In the case of \VOIII, the average value is $-4.0\pm60.6$, $-8.0\pm40.3$, and $0.0\pm23.2$ \kms, respectively for AGNs, composite objects,
and SF galaxies, indicating that the average \VOIII\ is similar among 3 groups. This is mainly due to the fact that the majority of 
objects have relatively low velocity shift, which can not be well constrained for given \VOIII\ uncertainties (see S.2.3). 
However, we clearly detect that the distribution of \VOIII\ of AGNs is a factor of $\sim$3 broader than that of SF galaxies,
suggesting the influence of gas outflows in AGNs. 

Considering the fact that galaxy bulge gravitational potential can also broaden the emission line, of which velocity dispersion is expected to correlate with stellar mass, we plot subsamples of AGNs and SF galaxies in the same stellar mass range, 10 $<$ log M$_{*}$ $<$ 10.5 (bottom panels in Figure 6). 
In this case, the range of velocity and velocity dispersion decrease, however the strong difference in the VVD distributions still remains. 
The mean \SOIII\ is $139.8\pm64.6$, $109.2\pm40.3$, and $83.1\pm24.7$, respectively for AGNs, composite objects, and SF galaxies,
showing a factor of $\sim$2 difference in terms of the mean value and dispersion of the distribution between AGNs and SF galaxies. 
The distribution of \VOIII\ is similar to that of the total sample, with the mean \VOIII\ = $-5.7\pm47.7$, $-6.1\pm31.8$, $-0.9\pm23.3$, respectively
for AGNs, composite objects, and SF galaxies.

\begin{figure*}
\centering
\includegraphics[width=.8\textwidth]{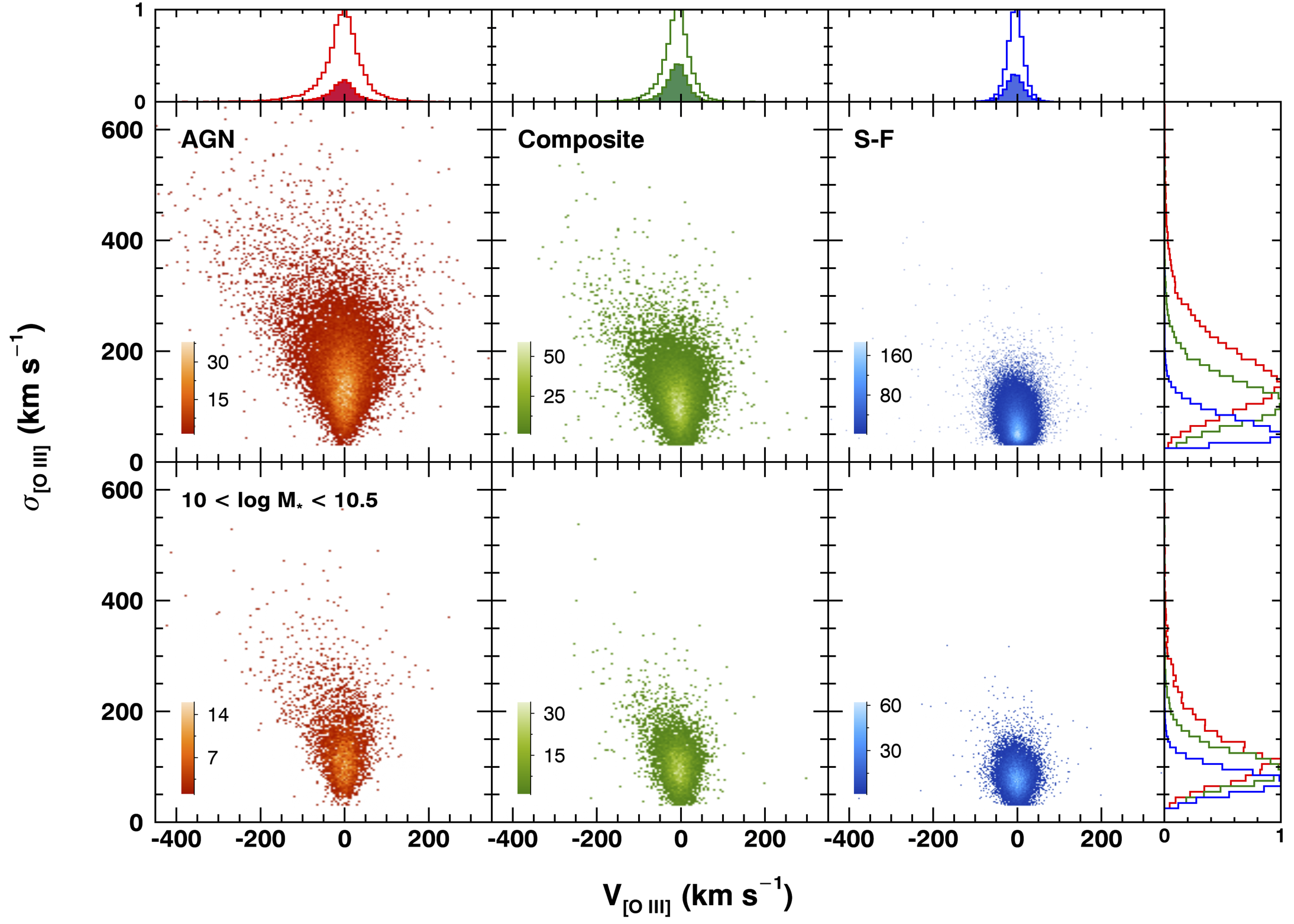} \\
\caption{\OIII\ Velocity-velocity dispersion diagrams, respectively for AGNs (left), composite objects (middle), and SF galaxies (right),
using the total sample (top) and the stellar mass-limited subsample (bottom). 
}
\end{figure*}

In Figure 7 we present the VVD diagram as a function of \OIII\ luminosity, respectively for AGNs (top), composite objects (middle), and SF galaxies (bottom).
In AGNs we find the VVD diagram dramatically expands toward higher values with increasing \OIII\ luminosity, indicating the gas outflows are driven by AGN energetics. Composite objects in general follow the same trend although at the highest luminosity bin, the number of objects is relatively small. 
In contrast, SF galaxies do not show any significant change of the VVD diagram as a function of \OIII\ luminosity, suggesting that photoionization mechanism 
of the \OIII\ line is not directly responsible for the \OIII\ kinematics. As we already showed in Figure 4, the gas kinematics are mainly governed by the gravitational potential of the host galaxy.

\begin{figure*}
\centering
\includegraphics[width=.9\textwidth]{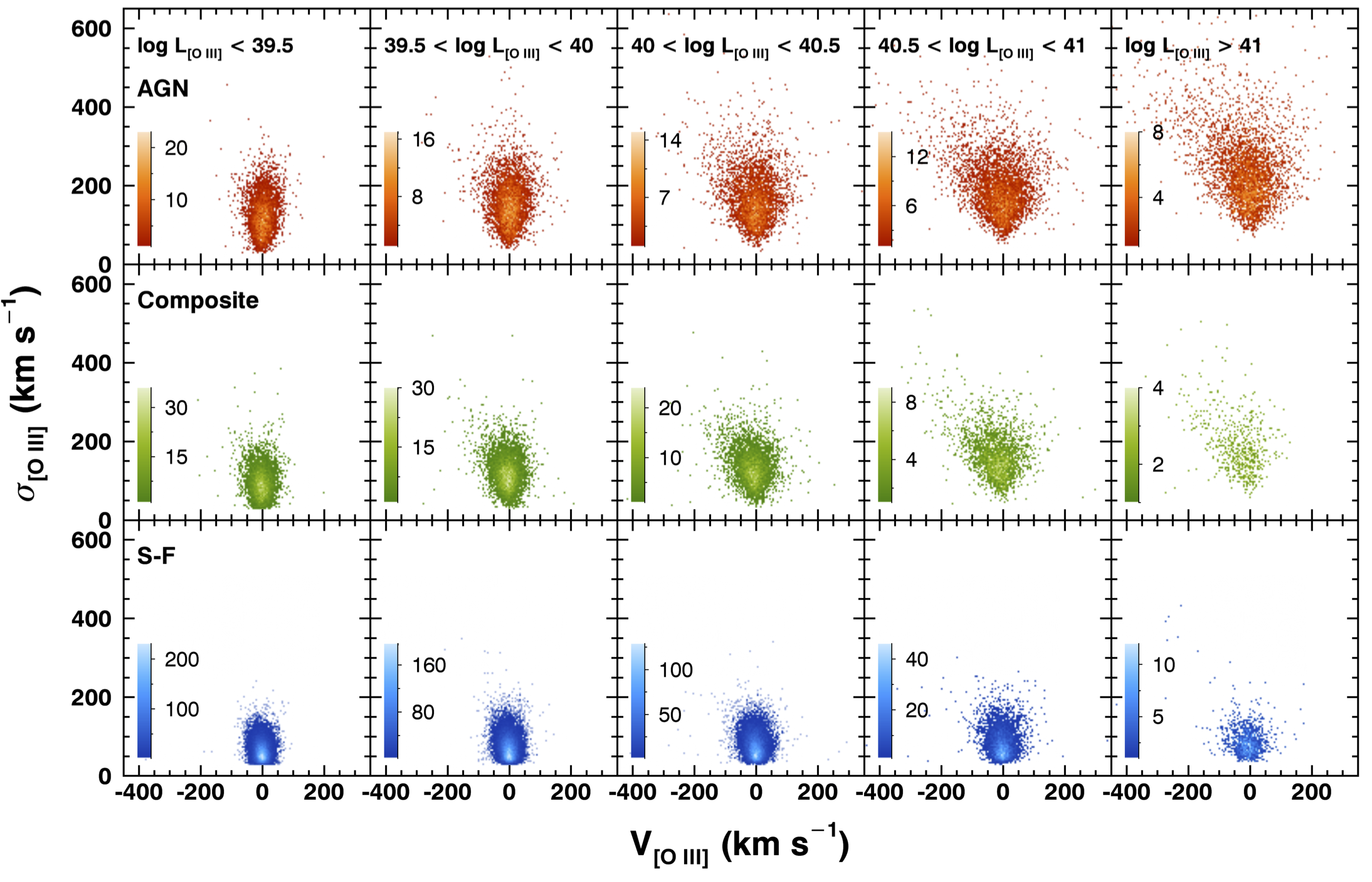} \\
\caption{\OIII\ Velocity-velocity dispersion diagrams as a fiction of \OIII\ luminosity, respectively for AGNs (top), composite objects (middle), and SF galaxies
(bottom). The color bar represents the number density calculated within a small bin with $\Delta$x= 5 \kms\ and $\Delta$y=4 \kms.}
\end{figure*}

If we investigate the VVD diagram as a function of stellar mass, instead of the \OIII\ luminosity, 
we find no dramatic expansion of the diagram. 
The VVD distribution is already broad even at low mass regime. Other than the increase of the sample size,  
AGNs, Composite objects and SF galaxies show similar distributions,
suggesting that stellar mass is not directly related to the expansion of the VVD diagram.

\begin{figure*}
\centering
\includegraphics[width=.8\textwidth]{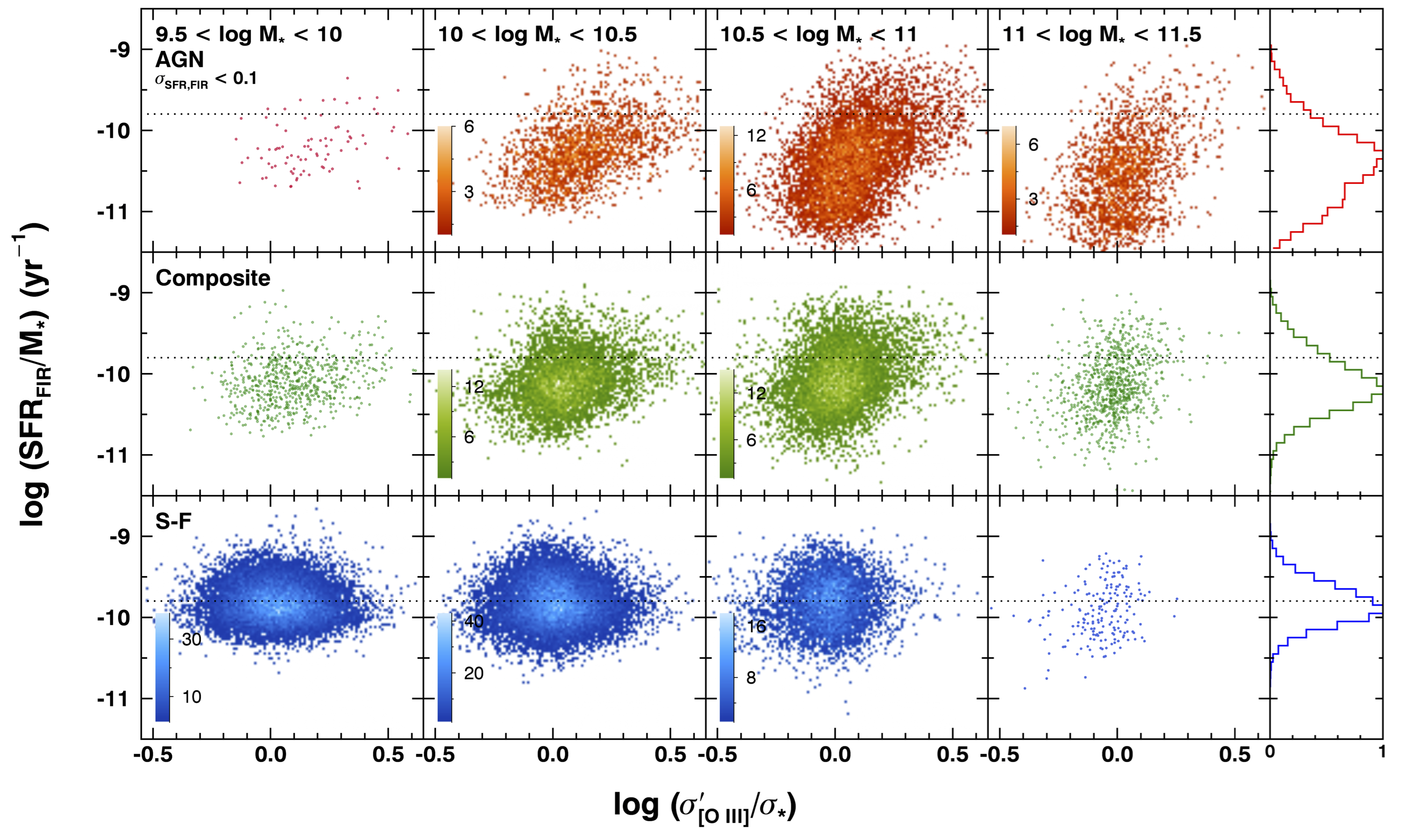} 
\includegraphics[width=.8\textwidth]{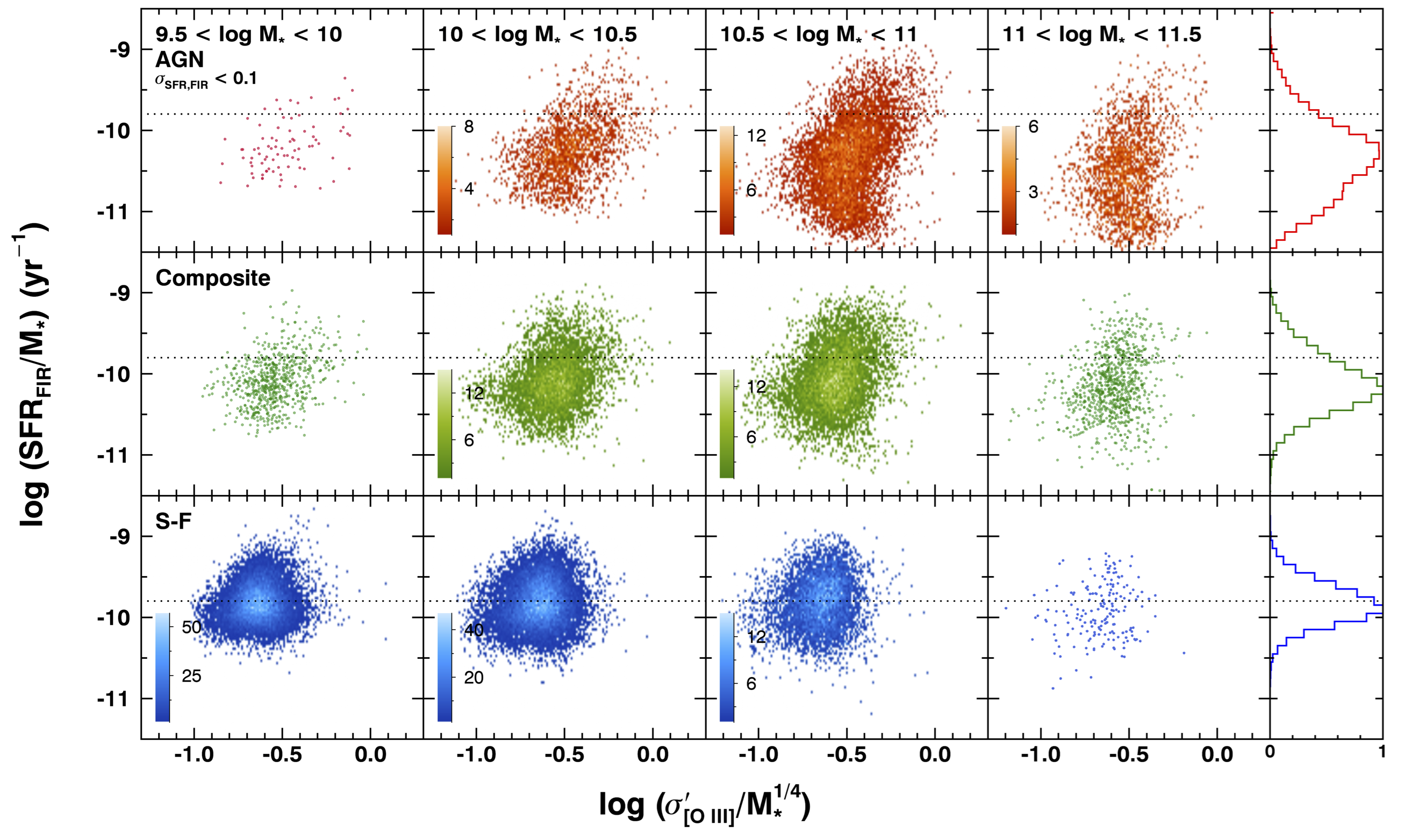} 
\caption{Top: SSFR vs. non-gravitational \OIII\ velocity dispersion for AGNs (top), composite objects (middle), and SF galaxies (bottom). 
The horizontal dotted line indicates the average SSFR of SF galaxies (log SSFR = $-$9.8).
Bottom: SSFR vs. non-gravitational \OIII\ velocity dispersion (\OIII\ divided by stellar mass)
The color bar represents the number density calculated within a small bin with $\Delta$x= 0.01 and $\Delta$y=0.03.
}
\end{figure*}

\subsection{Star formation vs. outflows}

\subsubsection{IR-based SFR}

In this section we investigate how the outflow kinematics are related to star formation in the host galaxies. 
Since the star formation rate (SFR) of AGN host galaxies is difficult to directly measure due to the contamination of AGN emission to SFR indicators, i.e., the H$\alpha$ line, UV continuum, etc \citep[e.g.,][]{Matsuoka15},
we use the SFR based on IR luminosity by adopting the catalogue of \citet{Ellison+16a}, which provides the IR luminosity of $\sim$330,000 SDSS galaxies and the estimated SFR converted from IR luminosity based on the conversion equation by \citet{Kennicutt1998},
\begin{equation}
\rm log L_{\rm IR} (\rm erg~ s^{-1}) = log SFR (M_{\odot} yr^{-1}) +43.591
\end{equation}
In their work \citet{Ellison+16a} determined IR luminosity from artificial neural network techniques, which were tested and trained based on the Herschel-SDSS Strip 82 matched sample of 1136 galaxies. As recommended by \citet{Ellison+16a}, we used reliable estimates of SFR using $\sigma_{\rm ANN}$ $<$ 0.1 \citep[see also][]{Ellison+16b}. In this case, a total of 30,352
objects (26.9\%) is excluded from our AGN+composite+SF sample in Table 1. More specifically, 10,316 objects from pure AGNs (45.3\%), 4,710 objects from composite objects (22.7\%), and 15,326 objects from SF galaxies (22.1\%) were removed due to the $\sigma_{\rm ANN}$ $<$ 0.1 criterion. Since we loose a large fraction of AGNs, we tried to use a less conservative criterion, i.e., 
$\sigma_{\rm ANN}$ $<$ 0.3, in order to include more objects with less secure SFR estimates. In this case, 12,042 objects (10.7\% of the total sample) were excluded (3,375, 1,781, and 6,886 objects respectively from pure AGNs, composite objects, and SF galaxies). The results based on the sample with $\sigma_{\rm ANN}$ $<$ 0.3 remained qualitatively the same. 
Note that the IR-based SFR represents the average SFR over $\sim$10$^{8}$ years as the IR emission is reradiated by cold dust around OB stars. 
In addition to IR-based SFR, we also used the D$_{4000}$ provided by the SDSS catalogue for a consistency check (see Section 4.3).

\subsubsection{Correlation between AGN outflows and SFR }

In Figure 8, we present the specific star formation rate (SSFR), by dividing the estimated SFR by stellar mass, as a function of the outflow velocities. To account for the difference of stellar mass range between AGN and SF galaxies, we present the comparison in each stellar mass bin. Here, the outflow velocities as expressed as $\sigma_{\rm [OIII]}'$, are calculated by adding the velocity and velocity dispersion of \OIII\ in quadrature, which can be understood as the corrected velocity dispersion of \OIII\ accounting for the projection effect \citep[for details, see][]{Bae&Woo16}. Then, the outflow velocities are normalized by stellar velocity dispersion in order to show the non-gravitational (i.e., outflow) kinematics. 

In the case of SF galaxies, the SSFR does not change along the stellar mass bin, showing that non-AGN galaxies form a star forming sequence with a constant SSFR within a factor of $\sim$2 scatter (see Table 2). In each stellar mass bin, the distribution of log SSFR is centered at the mean value of $-9.8\pm0.3$ yrs$^{-1}$, which is averaged over all SF galaxies. 
With respect to the normalized gas kinematics, we find no correlation, indicating that the kinematics of ionized gas are not related to SF. 
As expected, SFR is not affected by gas kinematics in SF galaxies since gas kinematics are mainly due to the galaxy gravitational potential rather than AGN outflows. 
Note that at lower stellar mass bin, stellar velocity dispersion is on average more uncertain, broadening the distribution of gas-to-stellar velocity dispersion ratio. 

In contrast, we find that AGNs and composite objects have on average lower SSFR than SF galaxies. 
The mean log SSFR is $-10.4\pm0.5$ and $-10.1\pm0.4$, respectively for pure AGNs and composite objects, 
showing a factor of 4 and 3 lower SFR than non-AGN galaxies at fixed stellar mass. 
Moreover, we find that the SSFR of AGN sample tend to increase with increasing outflow velocities. 
In the 10.5$<$ log M$_{\odot}$/M$_{\odot}$ $<$ 11 bin, for example, there is a clear increasing trend of SSFR with increasing outflow velocities in AGNs, while the distribution is almost round in SF galaxies. This trend implies a close connection between AGN outflows and star formation in the host galaxies. Interestingly, AGNs with higher outflow velocities have on average higher SSFR than AGNs with lower outflow velocities, implying that outflows do not instantly suppress SF, although the time scale of SF quenching has to be carefully investigated. Since AGNs with strong outflows show higher SFR than AGNs with weak outflows, this may indicate a close link between outflows and SF. 

We find no evidence of enhanced SF in AGNs. Even for the AGNs with extreme outflow velocities, SSFR is not very high and rather it is comparable to that of SF galaxies (dashed line), suggesting that host galaxies of extreme AGN outflows have regular SFR compared to SF galaxies. In contrast, SSFR decreases with decreasing outflow velocities.
For example, AGNs with no or weak outflows (i.e., when $\sigma^{'}_{\rm OIII} \sim \sigma_{*}$), SSFR is almost an order of
magnitude lower than SF galaxies or AGNs with strong outflows. In other words, AGNs with strong outflows
have similar SFR compared to SF galaxies, while AGNs with weak/no outflows have much lower SFR than non-AGN galaxies. Thus, the 0.6 dex mean difference of SSFR between AGN and SF galaxy samples is mainly due to the population of AGNs with weak \OIII\ outflows. 

The stellar velocity dispersion measured from SDSS spectra may not be the best parameter to represent the virial motion caused by the gravitational potential of the host galaxies, particularly for lower mass galaxies, since the spectral resolution of the SDSS spectroscopy is limited, and the effect of the rotational broadening and the projection due to the random inclination of the stellar disk to the line-of-sight can be significant for rotation-dominant galaxies.
Considering these effects, we use stellar mass, instead of \SVD, to normalize the gas kinematics.
Assuming the Fiber-Jackson relation, we use stellar mass to the 1/4 power (M$_*^{1/4}$) as a proxy for stellar velocity dispersion representing the virial motion (bottom panel in Fig. 9). We see qualitatively similar distribution of SSFR with the normalized outflow kinematics. Except for the AGNs with lowest SSFR, there is a general trend that AGNs with higher outflow velocities have higher SSFR, while the SSFR of AGNs with no outflows have much lower SSFR.

\subsubsection{Distribution of SSFR of AGNs and SF galaxies }

\begin{figure}
\centering
\includegraphics[width=.23\textwidth]{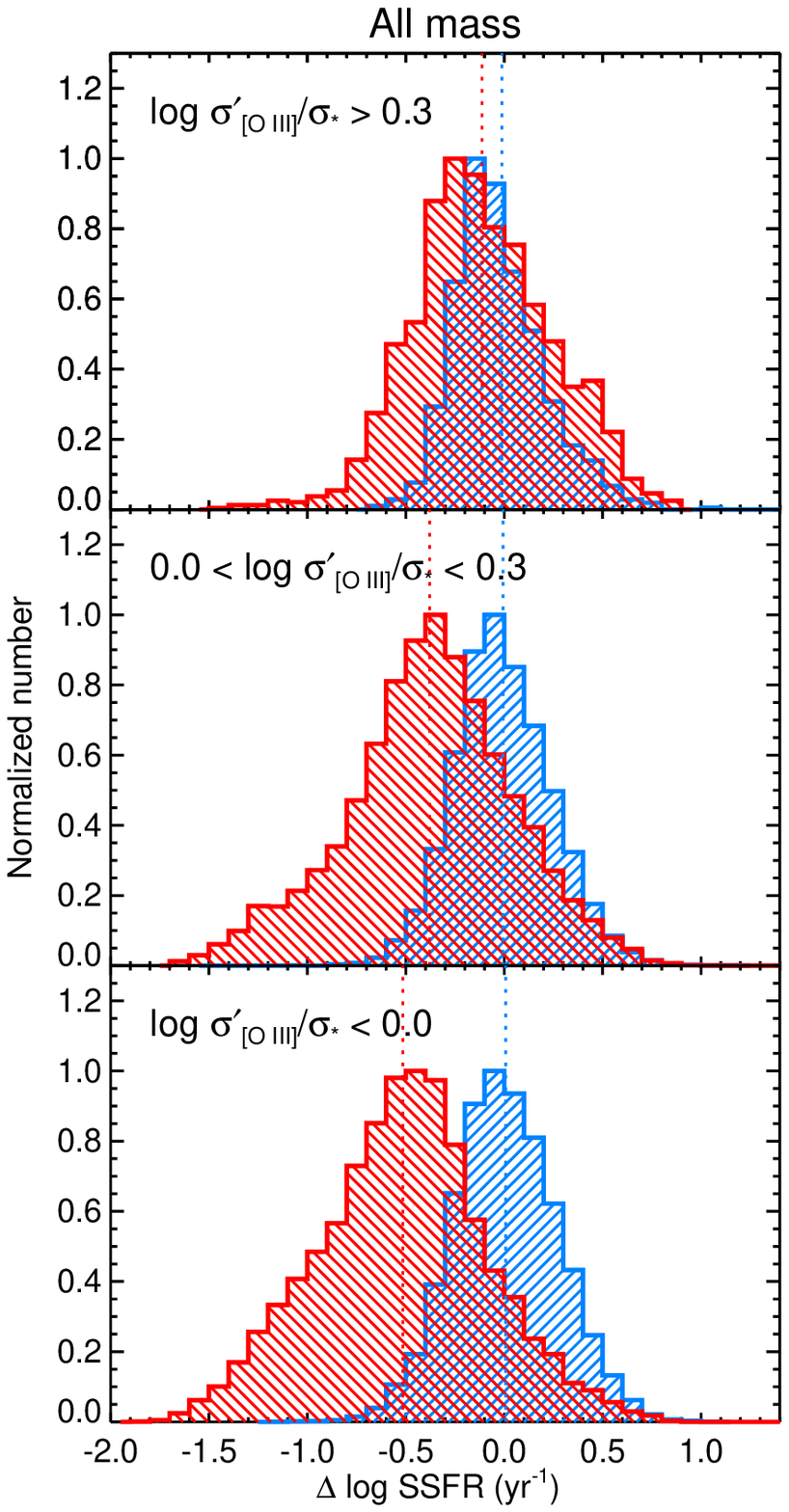}
\includegraphics[width=.23\textwidth]{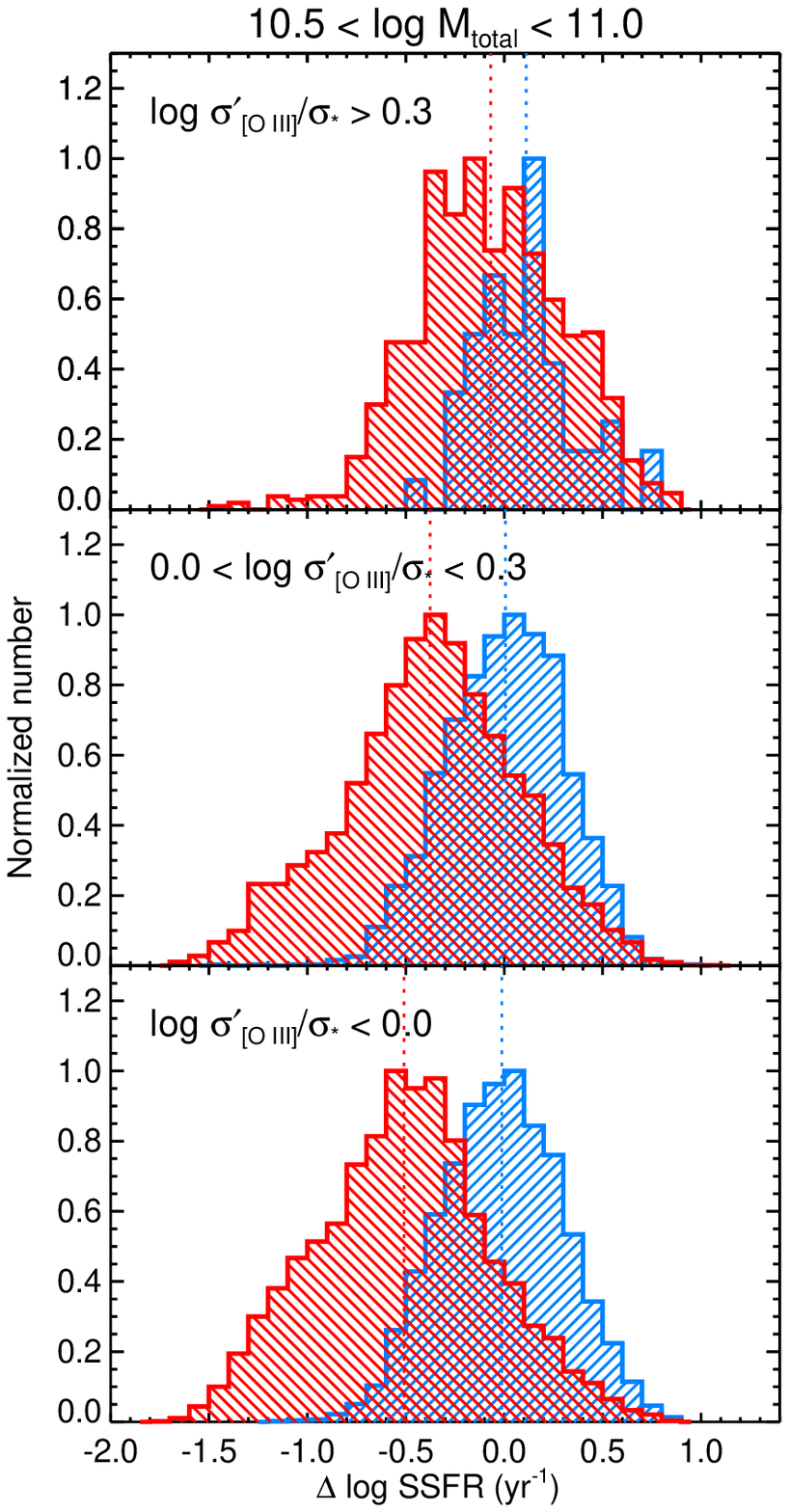}
\caption{Distribution of SSFR of AGN+composite objects (red) with respect to SF galaxies (red)
for all stellar masses (left), after divided into 3 groups; strong outflows (log $\sigma_{\rm OIII}'$/\SVD $>$ 0.3; top),
weak outflows (0 $<$ log $\sigma_{\rm OIII}'$/\SVD $<$ 0.3; middle), and no outflows (log $\sigma_{\rm OIII}'$/\SVD $<$ 0; bottom). In the left panels, AGNs and SF galaxies in a fixed stellar mass bin (10.5 $<$ log M$_{\odot}$ $<$ 11)
are presented. The relative SSFR is calculated with respect to the mean SSFR of all SF galaxies. 
}
\end{figure}

\begin{figure}
\centering
\includegraphics[width=.23\textwidth]{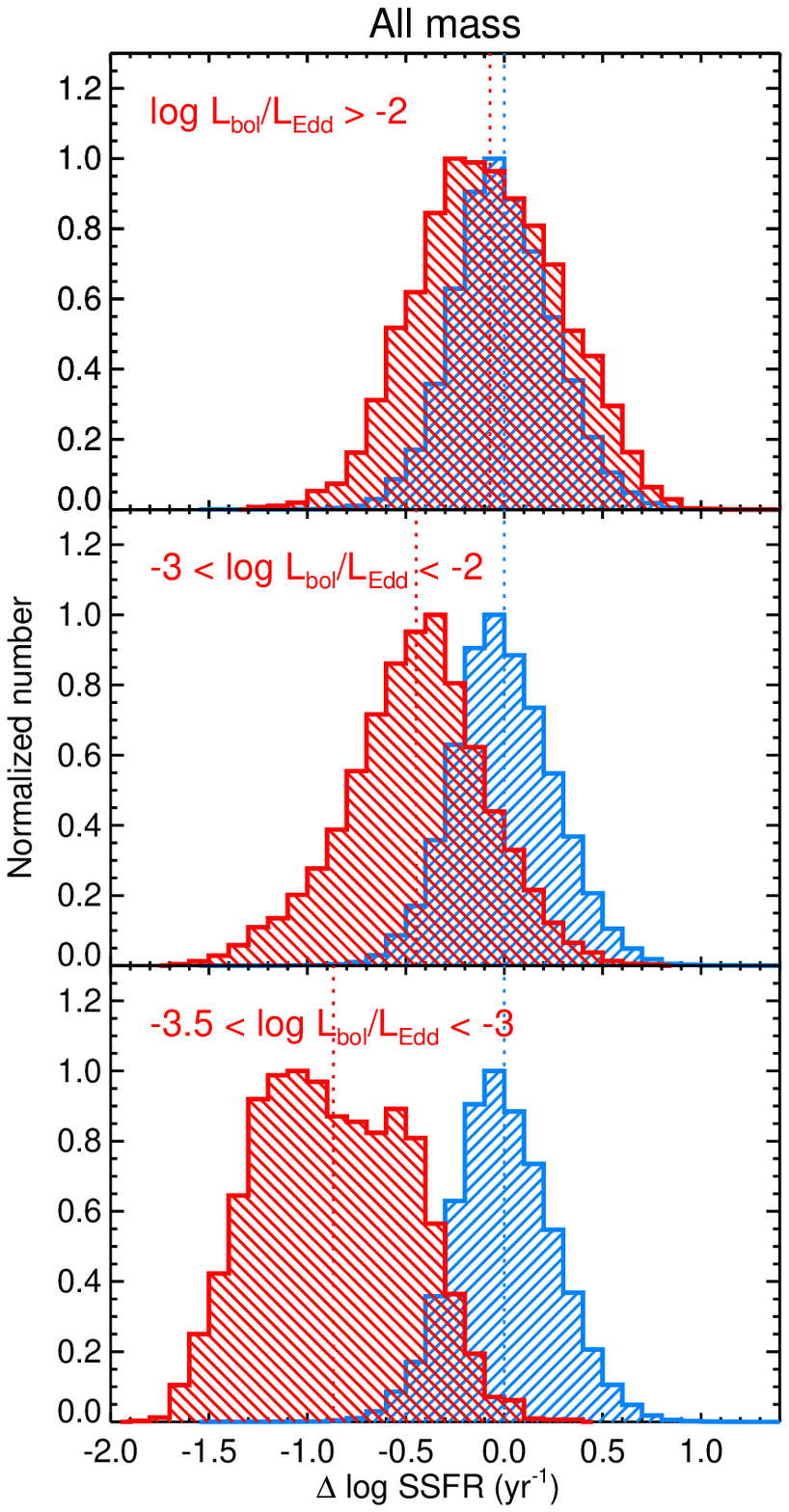}
\includegraphics[width=.23\textwidth]{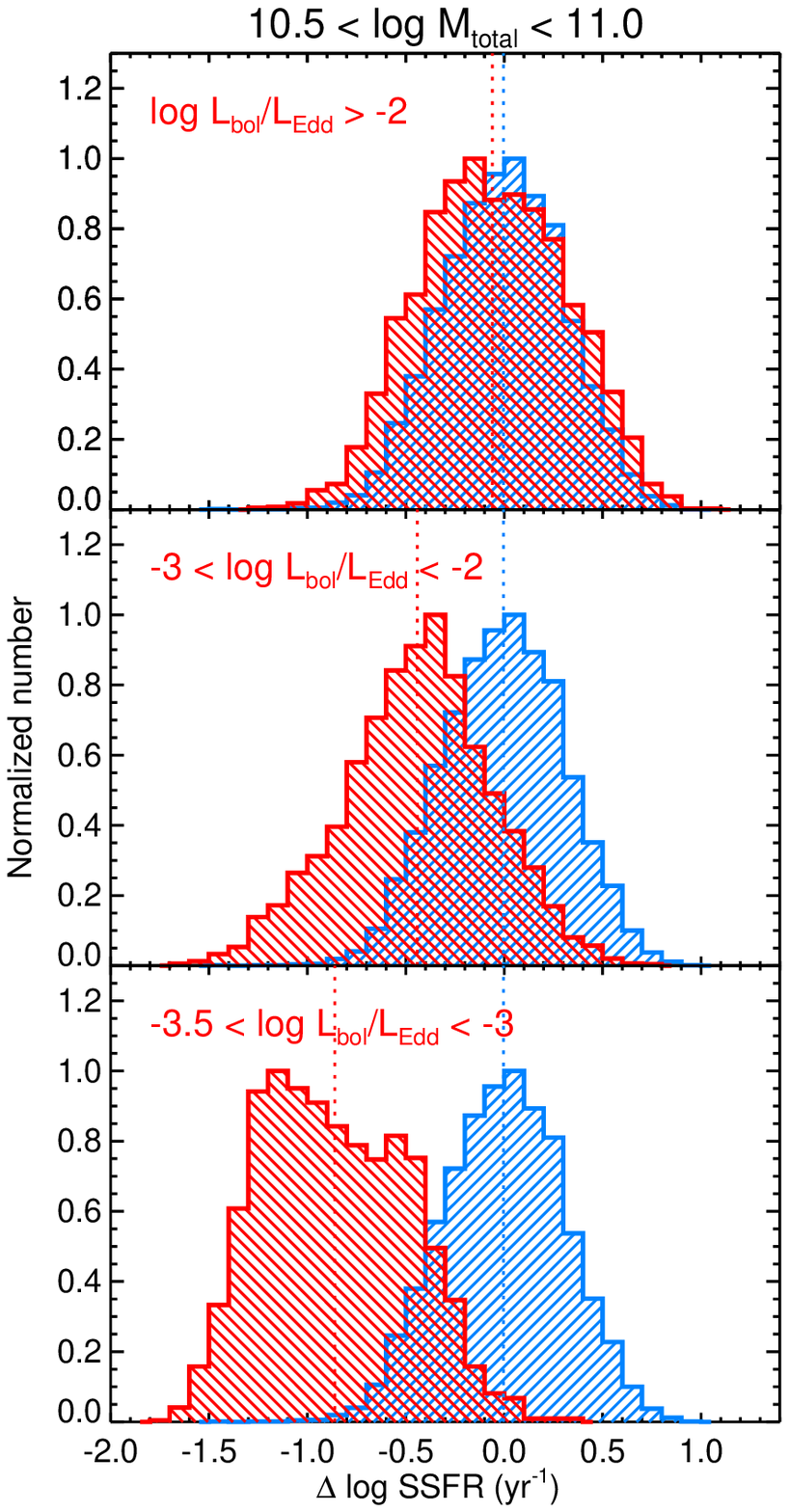}
\caption{Distribution of SSFR of AGN+composite objects (red) with respect to SF galaxies (red)
for all stellar masses (left), after divided into 3 Eddington ratio groups; high Eddington ratio (log L$_{bol}$/L$_{Edd}$  $>$-2; top), 
intermediate Eddington ratio (-3 $<$ log L$_{bol}$/L$_{Edd}$  $<$-2; middle), and low Eddington ratio (-3.5$<$ log L$_{bol}$/L$_{Edd}$  $<$-3; bottom).
In the left panels, AGNs and SF galaxies in a fixed stellar mass bin (10.5 $<$ log M$_{\odot}$ $<$ 11)
are presented. The relative SSFR is calculated with respect to the mean SSFR of all SF galaxies. 
Note that the Eddington ratio estimates based on the \OIII\ luminosity may have systematic uncertainties due to 
the contribution from the SF region to the observed \OIII\ luminosity.
}
\end{figure}

To further investigate the difference of SSFR between AGNs and SF galaxies, we compare the distribution of SSFR between AGNs (pure AGN + composite objects; red), and SF galaxies (blue) in Figure 9. Here we divide the sample into 3 groups: strong (i.e., log $\sigma_{\rm [OIII]}'$/\SVD $>$ 0.3; top panel), intermediate (i.e., 0$<$ log $\sigma_{\rm [OIII]}'$/\SVD$<$ 0.3; middle panel), and weak outflows (i.e., log $\sigma_{\rm [OIII]}'$/\SVD $<$0; bottom panel). While the distribution of SSFR is broad, AGNs with intermediate and weak outflows have lower SSFR by $0.38\pm0.43$ and $0.52\pm0.45$ dex, respectively, compared to the mean SSFR of SF galaxies. In contrast, AGNs with strong outflows have similar SSFR compared to SF galaxies (see Table 2 for details). 
Considering the stellar mass difference between AGNs and SF galaxy sample, we compare the SSFR distribution at fixed stellar bass bin, e.g., at $10.5<$ log M$_{*}$ $<11$ (right panel in Figure 9). We find a similar trend of decreasing mean SSFR with decreasing outflow velocities. 
Since the distribution of SSFR is very broad ($\sim$1 dex in FWHM), large ranges of AGN outflow velocities and SFR are present in the sample.
However, it is clear that there is strong trend of decreasing SSFR for AGNs with no outflows. 

Although it is previously found that optical and X-ray AGNs have on average lower SSFR than SF galaxies \citep[e.g.][]{Shimizu+15, Ellison+16b}, a physical mechanism for explaining the trend was not clearly identified. We try to interpret these results by connecting AGN outflows with SF in the following scenario. 
When gas is supplied, both AGN and SF are triggered as we typically observe the correlation between X-ray and IR luminosities \citep[e.g.][]{Netzer09}. 
However, it would take the order of a dynamical time for outflows to impact on the ISM over galactic scales. Once the outflows start playing a role in impacting on the ISM, we may observe the decrease of both SFR and AGN activity. After that, as the AGN activity decreases, the outflow kinematics become weaker and finally disappear. 
In this scenario, we link AGNs with strong outflows hosted by galaxies with regular SFR and AGNs with no outflows hosted by galaxies with lower SFR
as a evolutionary sequence. Perhaps, we are seeing for the first time, the physical link between AGN outflows and the delayed suppression of SF. 

We further investigate whether SSFR is related with AGN energetics
since outflow velocities and AGN luminosity show a broad correlation, albeit with a large scatter.
In Figure 10, we present the distribution of SSFR, after dividing AGNs into three Eddington ratio bins. 
Note that since we cannot detect a very weak \OIII\ line, the Eddington ratio of the AGN sample ranges from 10$^{-4}$ to 1 while most AGNs are
located within 10$^{-3.5}$ $<$ log L/L$_{Edd}$ $<$ -0.5 (see Figure 5 in Paper 1).
We find a similar trend that low Eddington AGNs have clearly lower SSFR than SF galaxies, while high Eddington AGNs have comparable SSFR with respect to SF galaxies. While the difference of SSFR distribution 
is negligible between high Eddington AGNs and SF galaxies with a small negative offset of $0.07\pm0.41$,
low Eddington AGNs clearly show much lower SSFR by $0.87\pm0.37$ dex, indicating that SFR is much lower in the 
host galaxies of low Eddington AGNs. The same trend is detected 
when we limit the sample within the stellar mass bin 10.5$<$ log M$_{*}$/M$_{\odot}$ $<$ 11 (right panel in Figure 10). 

Combing the trend of SSFR with outflow velocities and Eddington ratios, it seems that more energetic AGNs with strong
outflows and high Eddington accretion tend to be hosted by typical SF galaxies, while low-Eddington AGNs
with weak/no outflows have relatively quiescent host galaxies.

\begin{figure*}
\centering
\includegraphics[width=.8\textwidth]{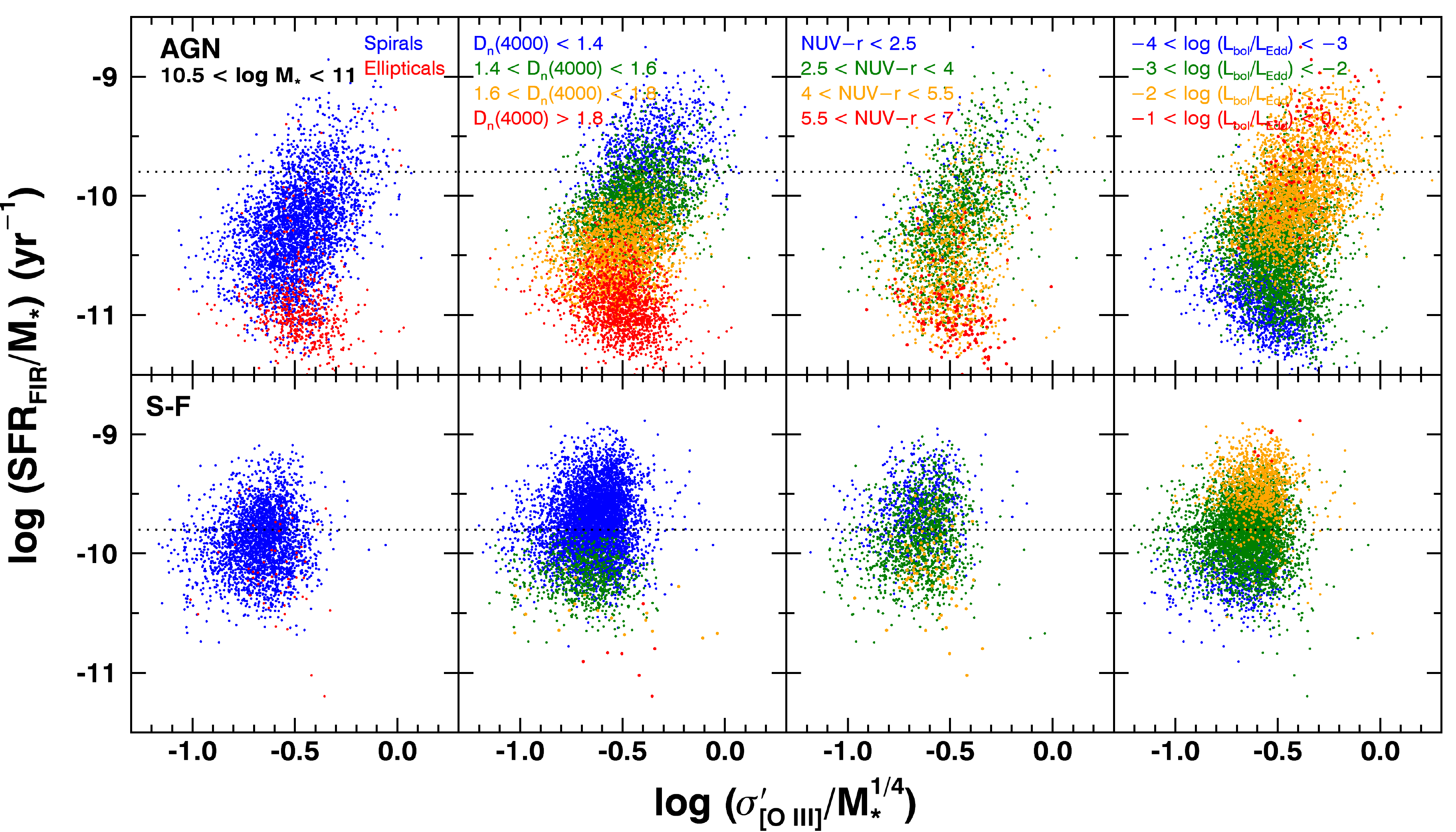} 
\caption{Comparison of SSFR and outflow kinematics normalized by stellar mass, respectively for pure AGNs (top)
and SF galaxies (bottom). Different colors represent galaxy morphology (1st col.), D4000 (2nd col.), NUV-r color (3rd col.),
and Eddington ratios (4th col.).
}
\end{figure*}

In Figure 11, we investigate how host galaxy properties are related with AGN outflows and SSFR. 
First, we adopt the galaxy morphology information from Galaxy Zoo \citep[1st col. in Figure 11;][]{Lintott11}. Since each galaxy is classified as elliptical, spiral, or uncertain object, we only present ellipticals and spirals by excluding uncertain objects. 
In AGN sample spiral galaxies generally follow the broad trend between SSFR and outflow velocities while elliptical galaxies are located separately at a much lower SSFR region. 
These elliptical galaxies would not be included in our emission line galaxy sample if AGN was not present since they have no or weak emission lines
from star forming region. However, by excluding these elliptical galaxies, we still see that the SSFR of AGNs ranges down to much lower level compared to that of SF galaxies. Considering the potential contamination 
of elliptical host galaxies in the AGN sample, we investigated the SSFR distribution using only spiral galaxies. We find that
the difference of mean SSFR between AGNs and SF galaxies and the decreasing SSFR with decreasing outflow velocities
and Eddington ratios remain the same, indicating that the contamination of elliptical galaxies is not mainly responsible for the much lower 
SSFR of AGNs with no outflows. 
Second, we color-code AGNs based on the 4000\AA-break strength D$_{4000}$ as a SFR indicator \citep{Brinchmann2004}, which is adopted from the MPA-JHU catalogue
(2nd col. in Figure 11). As expected SSFR correlates with D$_{4000}$ in AGNs, while SF galaxies have overall lower values of D$_{4000}$. 
The decreasing trend of D$_{4000}$ with increasing outflow velocities confirms a consistency between IR-based SFR and  D$_{4000}$-based SFR. 
Third, we also utilize the UV-to-optical color as a SFR indicator. By matching our SDSS sample with GALEX General Release
6/7 including AIS and MIS data \citep{Morrissey07}, we obtained NUV-r color for 38,105 objects. As shown in Figure 11 (3rd col.),
we find qualitatively the same trend although the size of the matched sample is relatively small. 
Finally, we present Eddington ratios in the 4th col. in Figure 11, showing a clear increasing trend with both SSFR and outflow velocities,
albeit with considerable scatter. These findings indicate that a clear trend, although not bimodal, is present between high Eddington, strong outflow AGNs in SF galaxies and low Eddington, no outflow AGNs in weakly star-forming galaxies.

\begin{table*}
\begin{center}
\caption{$\Delta$ SSFR}
\begin{tabular}{ccccc}
\tableline\tableline
  & AGN+composite & AGNs & Composite  & S-F \\
\tableline
&& all mass \\
(1)&(2)&(3)&(4)&(5)\\
\tableline
\tableline
0.3 $<$ log $\sigma_{\rm OIII}$'/\SVD            & $-0.11\pm0.35$ &    $-0.15\pm0.37$ & $-0.05\pm0.32$ & $-0.01\pm0.43$\\
0 $<$ log $\sigma_{\rm OIII}$'/\SVD $<$ 0.3 & $-0.38\pm0.43$ &     $-0.54\pm0.45$ & $-0.24\pm0.36$ & $-0.01\pm0.25$\\
log $\sigma_{\rm OIII}$'/\SVD $<$ 0              &  $-0.52\pm0.45$ &   $-0.78\pm0.40$ & $-0.35\pm0.39$ & $0.01\pm0.26$\\
\tableline
-2 $<$log L$_{bol}$/L$_{Edd}$             &  $-0.07\pm0.41$ & $-0.19\pm0.35$ & $     0.09 \pm 0.43$ &                            \\
-3 $<$ log L$_{bol}$/L$_{Edd}$ $<$ -2 &  $-0.45\pm0.36$ & $-0.66 \pm0.35$ & $     -0.32 \pm0.29$ & $    0.00 \pm0.26$\\
log L$_{bol}$/L$_{Edd}$ $<$ -3             & $-0.87\pm0.37$ & $-1.12 \pm 0.26$ & $     -0.59 \pm0.27$ &                           \\
\tableline
\tableline

&& 10.5 $<$ log M$_{\odot}$ $<$ 11 \\
\tableline
0.3 $<$ log $\sigma_{\rm OIII}$'/\SVD              & $-0.07\pm0.37$ &    $-0.12\pm0.37$ & $0.05\pm0.34$ & $0.11\pm0.25$\\
0 $<$ log $\sigma_{\rm OIII}$'/\SVD $<$ 0.3 & $-0.38\pm0.44$ &     $-0.53\pm0.45$ & $-0.21\pm0.36$ & $0.01\pm0.30$\\
log $\sigma_{\rm OIII}$'/\SVD $<$ 0              &  $-0.51\pm0.44$ &   $-0.76 \pm0.41$ & $-0.33\pm0.37$ & $-0.01\pm0.31$\\
\tableline
-2 $<$log L$_{bol}$/L$_{Edd}$             &  $-0.06\pm0.37$ & $-0.18\pm0.35$ & $     0.17 \pm 0.29$ &   \\
-3 $<$ log L$_{bol}$/L$_{Edd}$ $<$ -2 &  $-0.44\pm0.38$ & $-0.66 \pm0.35$ & $     -0.29 \pm0.31$ & $    0.00 \pm0.30$\\
log L$_{bol}$/L$_{Edd}$ $<$ -3             & $-0.86\pm0.36$ & $-1.10 \pm 0.25$ & $     -0.59 \pm0.27$ &  \\
\tableline
\tableline
&& 10 $<$ log M$_{\odot}$ $<$ 10.5 \\
\tableline
0.3 $<$ log $\sigma_{\rm OIII}$'/\SVD              & $-0.15\pm0.32$ &    $-0.21\pm0.32$ & $-0.10\pm0.30$ & $-0.02\pm0.23$\\
0 $<$ log $\sigma_{\rm OIII}$'/\SVD $<$ 0.3 & $-0.32\pm0.33$ &     $-0.45\pm0.35$ & $-0.27\pm0.31$ & $-0.02\pm0.25$\\
log $\sigma_{\rm OIII}$'/\SVD $<$ 0              &  $-0.43\pm0.34$ &   $-0.65 \pm0.31$ & $-0.37\pm0.32$ & $-0.01\pm0.27$\\
\tableline
-2 $<$log L$_{bol}$/L$_{Edd}$             &  $-0.06\pm0.32$ & $-0.20\pm0.30$ & $     0.04 \pm 0.31$ &                       \\
-3 $<$ log L$_{bol}$/L$_{Edd}$ $<$ -2 &  $-0.41\pm0.28$ & $-0.60 \pm0.28$ & $     -0.36 \pm0.26$ & $-0.02 \pm0.26$\\
log L$_{bol}$/L$_{Edd}$ $<$ -3             & $-0.65\pm0.28$ & $-0.90 \pm 0.22$ & $     -0.56 \pm0.24$ &                       \\
\tableline
\tableline
\end{tabular}
\tablecomments{(1) Range (2) mean $\Delta$SSFR for AGN+composite obejcts; (3) mean $\Delta$SSFR for AGNs;
(4) mean $\Delta$SSFR for composite objects; (5) mean $\Delta$SSFR for SF galaxies.}
\end{center}
\end{table*}

\section{Discussion}

\subsection{Delayed feedback or gas depletion?}

We reported that optical type 2 AGNs have on average lower SSFR compared to SF galaxies as previously found 
by \citet{Shimizu+15} and \citet{Ellison+16b} based on X-ray, optical, and radio AGNs. 
However, there is a large range of SSFR among AGNs, and we discovered a link between AGN outflows and SSFR.
While AGNs with powerful outflows have comparable SSFR with respect to SF galaxies,
AGNs with weak or no outflows tend to have much lower SSFR.
Moreover, there is a clear trend that the mean SSFR decreases with decreasing Eddington ratios. 
Combining these two trends, we see a distribution from AGNs with strong outflow, high Eddington accretion, and regular SFR
to AGNs with no outflow, low Eddington accretion, and much lower SFR. 
The trend between Eddington ratio and SSFR naturally explains the observed AGN-star formation luminosity relation in the local
universe \citep{Netzer09, Chen+13, Matsuoka15}. 

We may interpret these trends as a evolutionary sequence i.e., when there is gas supply, both AGN and SF are triggered. 
AGNs with high-Eddington accretion develop strong outflows while SF is on-going as in regular SF galaxies.
When the outflows start impacting on the ISM after the order of a dynamical time, the effect of AGN feedback is observable as the SFR decreases
while AGN becomes weaker and no strong outflows are visible. Note that the SFR based on IR luminosity traces the reradiation by cold dust around OB stars and averaged over $\sim$10$^{8}$ yrs. Also, the Eddington ratio was calculated based on \OIII\ luminosity, 
which does not reflect the short time scale 
flickering of accretion disk activity. Rather, \OIII\ luminosity is averaged over
longer time scale. 
If the outflow can effectively push the gas supply and/or prevent the cooling of the ISM, it is naturally expected that both SFR and AGN accretion will decrease. The distribution of AGNs from strong outflow, high Eddington, and regular SFR to no outflow, low Eddington, and low SFR
may be understood as the evolutionary sequence caused by delayed AGN feedback. 

Alternatively, the same evolutionary sequence can be interpreted as the consequence of depletion of gas supply.
When gas is supplied, both AGN and SF are triggered. However, once gas is depleted, both AGN activity and SF decreases.
Thus, we naturally expect that AGNs with strong outflow, high Eddington, and regular SFR become AGNs with no outflow, low Eddington,
and low SFR. In this case, no AGN feedback is invoked and simply gas depletion causes the transition. 

Instead of evolutionary sequence, it is also possible that intrinsic gas content varies among galaxies at given stellar mass. 
If the amount of gas is intrinsically large, AGNs may tend to be high-Eddington with strong outflows and host galaxies show regular
SFR as in SF galaxies. On the other hand, if gas content is intrinsically lower, SFR is lower than that of galaxies in star-forming sequence,
and AGNs have relatively low Eddington and weak outflows. 

For given our data set, it is not clear which scenario best explains the observed trend. In particular, in the evolutionary sequence scenario,
it is very difficult to confirm whether AGN feedback causes the transition or the transition is natural outcome of gas depletion. 
In the case of the 3rd scenario, we may be able to investigate the difference of intrinsic gas content based on CO observations
by selecting AGNs with strong and no outflow for given stellar mass and morphology.

\subsection{Bias against high star-formation rate galaxies?}

The decreasing trend of SSFR with decreasing Eddington ratio down to 10$^{-3.5}$ shown in Figure 10
can be interpreted in two different ways: selection bias and intrinsic nature.
We first examine whether the lower SSFR of low Eddington AGNs is due to the selection bias. 
Since we classify AGNs based on the emission line flux ratios in the BPT diagram, if AGNs produce weak emission lines, which can be over-shined by the emission produced by strong star formation, the combined observed emission lines are likely to be classified as SF galaxies rather than AGNs. 
Thus, it is possible that we may miss very weak black hole activity present in strongly SF galaxies due to the limitation of the optical BPT classification. 
Note that we do not consider very weak black hole activity with extremely low Eddington ratios ($<<$10$^{-4}$) 
\citep[e.g.,][]{Gallo+2008, Miller+2015}
since these very weak activities are not generally considered as AGN population and the feedback energy from these objects is expected
to be negligible in the context of galaxy evolution. 
If we focus on low Eddington ratio AGNs between 10$^{-4}$ and 10$^{-2}$, it is not clear why we should expect these AGNs
are dominantly hosted by highly SF galaxies, while we expect that very weak black hole activity is present in most galaxies. 
By selecting SF galaxies, \cite{Chen+13} reported that the mean SF luminosity correlates with the mean AGN X-ray luminosity 
down to Eddington ratio $\sim10^{-3}$ \citep[see also][]{Rafferty+11}, 
although it is extremely difficult to investigate the distribution of Eddington ratios below $\sim10^{-3}$ \citep[see e.g.][]{Jones+16}. 

For investigating whether our optical AGN sample misses low Eddington AGNs hosted by strongly SF galaxies,
X-ray AGN sample is very useful since optical BPT-based classification can be avoided. In fact, \citet{Shimizu+15} reported that
their Swift-BAT AGN sample shows on average lower SSFR compared to SF galaxies, indicating the same trend as we found 
in our sample. 

We independently investigate this issue by selecting X-ray AGNs from Swift-BAT 70 month catalogue \citep{Baumgartner+13} after matching with our SDSS sample.
Although the number of X-ray AGNs, for which we can determine stellar mass and Eddington ratio, 
is relatively small, these 59 X-ray AGNs show a consistent trend that AGNs with low Eddington ratio tend to show lower SSFR than AGNs with high Eddington ratio. 
For example, since the Eddington ratio (log L$_{bol}$/L$_{Edd}$) of this matched sample ranges from -3 to 0,
we divide them into 3 groups. We find that the D$_{4000}$-based SSFR of high Eddington AGNs (log L$_{\rm bol}$/L$_{\rm Edd}$ $>$ -1) is higher by 0.13 and 0.83 dex, respectively, than that of intermediate (-2$<$ log L$_{\rm bol}$/L$_{\rm Edd}$ $<$-1 ) and low Eddington AGNs (log L$_{\rm bol}$/L$_{\rm Edd}$ $<$ -2), although the distribution of SSFR in each group is very large with rms of 0.7 to 0.8 dex
due to presumably small number statistics. 
Note that we were not able to use IR-based SFR since IR luminosity taken from \citet{Ellison+16a} is available only 
for 20 out of 59 objects. 
Although we see a consistent trend between Eddington ratios and SSFR based on the X-ray sample, we cannot firmly conclude whether the lack of low-Eddington AGNs in SF galaxies in our optical sample is due to selection effect or intrinsic nature
since the size of the matched sample is very small and the X-ray flux limit of the Swift-BAT sample is still shallow (down to L$_{\rm bol}$/L$_{\rm Edd}$ $<$$\sim$-3). 

In investigating the trend between SSFR and Eddington ratio, it is important to reliably estimate AGN bolometric luminosity.
In the case of optically-identified type 2 AGNs, the \OIII\ luminosity is typically used for calculating Eddington ratios as in our study \citep[e.g.][]{ka03, Choi+2009}. However, the bolometric correction of \OIII\ is difficult to determine due to various effects, including dust extinction and the dependency on the ionization parameter \citep{Netzer09, Matsuoka15}. Also, star forming region can contribute to the \OIII\ flux observed through an aperture larger than the size of the narrow-line region. Therefore, it is important 
to decompose the AGN and SF fraction in the observed \OIII\ luminosity \citep[e.g.,][]{Jones+16}. By simply assuming the maximum AGN fraction in the \OIII\
luminosity of SF galaxies as 1\%, 10\%, and 50\%, we simulate the distribution of SSFR as in Figure 10 in order to test the effect of the contribution of non-detected AGNs in SF galaxies. We find that the case of 
1\% and 10\% AGN luminosity does not change the decreasing trend of the SSFR as a function of Eddington ratios down to 
10$^{-3}$. When we assume 50\% of \OIII\ in SF galaxies is originated from AGNs, we still see the same trend while the difference of
SSFR between high and low Eddington AGNs becomes smaller. 
 
Due to the aforementioned issues and uncertainty of Eddington ratio, it is clear that more detailed investigations are 
required to confirm the trend of the decreasing SSFR with decreasing Eddington ratios. 
In particular, a detailed comparison between X-ray and optical AGNs, for example, using the optical follow-up survey of the Swift-BAT AGNs \citep{Koss+2017}, will be very useful to constrain the nature of host galaxies of low Eddington AGNs, which is beyond the scope of this paper. 

\subsection{Uncertainty of IR-based SFR}

Since we have adopted the estimated SFR based on IR from \citet{Ellison+16a}, it is important to discuss whether the main results 
of this paper is affected by the uncertainty of IR-based SFR. 
As shown In Figure 11, other SFR indicators, i.e., D$_{4000}$ and UV-to-optical color show consistent results.
While SF galaxies have overall lower values of D$_{n}$4000, there is a clear spread of D$_{4000}$ among AGNs, which broadly correlates with outflow velocities.
Also, the UV-to-optical color shows a similar trend with AGN outflow velocities.
These results suggest that although the IR-based SFR is not the best SF indicator, there is a clear trend of SSFR with AGN outflow velocities,
while SF galaxies do not show such a trend.

\section{Summary \& Conclusion}

Using a large sample of $\sim$110,000 type 2 AGNs out to z$\sim$0.3, we investigated the kinematics of the ionized gas outflows, demography of
outflows in AGNs and SF galaxies, and the connection between outflows and SF. 
We summarize the main results as follows.

\medskip

$\bullet$ Gas and stellar velocity dispersions are comparable to each other in SF galaxies, indicating that he gravitational potential of host galaxies
determines are mainly responsible for gas and stellar kinematics. 

$\bullet$ In contrast,  AGNs show much larger \OIII\ velocity dispersion
than stellar velocity dispersion, indicating that strong non-gravitational kinematics, i.e., outflows, are present. The kinematic component of outflows 
is comparable to or stronger than the virial motion caused by the gravitational potential. 

$\bullet$ The fraction of AGNs with outflows steeply increases with AGN luminosity and Eddington ratio. In particular, the majority of luminous AGNs presents strong non-gravitational kinematics in the \OIII\ profile. 

$\bullet$ We find a dramatic difference of the outflow signatures between AGNs and star-forming galaxies.
The distribution in the \OIII\ velocity - velocity dispersion diagram dramatically expands toward large values with increasing AGN luminosity, implying that the outflows are AGN-driven, while that of SF galaxies show no significant change as a function of \OIII\ luminosity.

$\bullet$  The SSFR of non-outflow AGNs is much lower than that of strong outflow AGNs, while the SSFR of strong outflow AGNs is comparable to that of SF galaxies.
We interpret this trend as a result of delayed AGN feedback as it takes the order of a dynamical time for the outflows to suppress star formation. Alternatively, gas depletion or
intrinsic difference of gas content may cause the trend.

\acknowledgments

We thank the anonymous referee for the suggestions, which were useful to clarify the results and interpretation. 
We thank Sara Ellison for her helpful comments on the IR-based SFR. 
Support for this work was provided by the National Research Foundation of Korea grant funded by the Korea government (No. 2016R1A2B3011457).

\bibliographystyle{apj}

%http://merkel.zoneo.net/Latex/natbib.php
%\citet{jon90}	    		-->    	Jones et al. (1990)
%\citet[chap. 2]{jon90}	   	-->    	Jones et al. (1990, chap. 2)
%\citep{jon90}	   		-->    	(Jones et al., 1990)
%\citep[chap. 2]{jon90]	-->    	(Jones et al., 1990, chap. 2)
%\citep[see][]{jon90}	   	-->    	(see Jones et al., 1990)
%\citep[see][chap. 2]{jon90}	-->    	(see Jones et al., 1990, chap. 2)
%\citet*{jon90}	    		-->    	Jones, Baker, and Williams (1990)
%\citep*{jon90}	    		-->    	(Jones, Baker, and Williams, 1990)
%\citet{jon90,jam91}	    	-->    	Jones et al. (1990); James et al. (1991)
%\citep{jon90,jam91}	    	-->    	(Jones et al., 1990; James et al. 1991)
%\citep{jon90,jon91}	    	-->    	(Jones et al., 1990, 1991)
%\citep{jon90a,jon90b} 	-->    	(Jones et al., 1990a,b)

\end{document}